\UseRawInputEncoding
\documentclass{jfm}
\usepackage{graphicx}
\usepackage{epstopdf, epsfig}
\usepackage{subfig}
\usepackage{epsf}
\usepackage{color}
\usepackage{colortbl}
\usepackage{array}
\usepackage{booktabs}
\usepackage{color}
\usepackage{bm}
\usepackage{hyperref}
\usepackage{natbib}
\usepackage{amsmath}
\usepackage{array}
\usepackage{upgreek}
\usepackage[dvipsnames]{xcolor}
\hypersetup{
    colorlinks = black,
    urlcolor   = blue,
    citecolor  =blue,
}

\newcommand{\RomanNumeralCaps}[1]
\linenumbers
\newcommand{\comm}[1]{}

\shorttitle{Wake instability past a sphere settling in a strongly stratified flow}
\shortauthor{C.~F.~Mo et al.}

\title{Wake instability past a sphere settling in a strongly stratified fluid}

\author{Chang-Fan Mo\aff{1}, Matthieu J. Mercier\aff{2}, Jacques Magnaudet\aff{2} \and Jie Zhang\aff{1}\corresp{\email{j\_zhang@xjtu.edu.cn}}}

\affiliation{\aff{1}State Key Laboratory for Strength and Vibration of Mechanical Structures, School of Aerospace, Xi'an Jiaotong University, Xi'an, Shaanxi, PR China
\aff{2}Univ. Toulouse, Toulouse INP, CNRS, IMFT (Institut de M\'ecanique des Fluides de Toulouse), Toulouse, France}

\begin{document}
\maketitle

\begin{abstract}
The wake of a body moving across the isopycnals of a strongly stratified fluid is characterized by the presence of an intense jet which, under certain circumstances, may become unstable. 
To get insight into the phenomenology of this instability and the underlying mechanisms, we conduct fully-resolved three-dimensional time-dependent simulations of the flow past a rigid sphere settling steadily through a linearly stratified fluid 
over a wide range of flow parameters which include the case of salt-stratified water.
Results reveal a rich dynamics characterized by distinct wake symmetries, vortical structures and transverse force signatures. Simulations evidence the existence of varicose and sinuous instability modes that arise from distinct physical mechanisms. Thanks to several metrics, we build a phase map delineating the bounds of each instability regime as a function of the three control parameters of the problem, namely the Froude, Reynolds and Prandtl numbers. 
We show that the instability mechanism results from a subtle interplay between baroclinic vorticity generation, vortex tilting and radial transport of the mean density gradient within the jet. 

\end{abstract}


\section{Introduction}\label{sec1}

Particles and microorganisms moving through the atmosphere and oceans are confronted with stratified environments in which the fluid density varies with depth \citep{bergstrom1997behavioural,widder1999thin,sutor2008effects,ardekani2017transport}. Density gradients, typically induced by temperature or salinity, impose specific effects that can slow down or even arrest the otherwise straightforward settling of nearly neutrally-buoyant particles such as that of marine snow in the ocean \citep{alldredge2002occurrence,stocker2012marine,prairie2013delayed,prairie2015delayed,auta2017distribution,ahmerkamp2022settling}. Understanding and accurately modeling the dynamics of particles settling in stratified fluids is therefore essential for advancing our knowledge of oceanic biogeochemical cycles \citep{denman1995biological,prairie2015delayed}, constraining mechanisms of climate variability \citep{munk1966abyssal}, and mitigating observational biases in environmental measurement systems \citep{bewley2016efficient}. As a simplified, yet canonical, model, the motion of a single sphere of radius $R$ settling at constant speed $U$ through a linearly stratified fluid (characterized by a Brunt-V\"ais\"al\"a  frequency, $N$, kinematic viscosity, $\nu$, and molecular diffusivity of the stratifying agent, $\kappa$) offers valuable insights into fluid-solid coupling issues under stratified conditions \citep{magnaudet2020particles,more2020motion}, and contributes to improve the understanding of mechanisms underlying particulate transport in more complex stratified flows.

Depending on the governing dimensionless parameters characterizing the settling body and the surrounding stratification, laboratory experiments have revealed a variety of wake regimes behind a sphere descending vertically in salt-stratified water \citep{hanazaki2009jets}, where the Prandtl number, $Pr=\nu/\kappa$, is approximately $700$. In particular, beyond a critical density gradient - quantified by the Froude number $Fr=U/NR$ - the classical vortex ring at the back of the body collapses and reorganizes itself into a vertically aligned jet in which the fluid may exhibit large upward velocities. This transition is driven by buoyancy forces resulting from the entrainment and subsequent upward displacement by the settling sphere of fluid parcels heavier than those in its wake. This jet structure was reproduced in several numerical studies \citep{torres2000flow,hanazaki2009schmidt,hanazaki2015numerical}, which further demonstrated that its formation and geometry are strongly influenced by both stratification and density diffusion. Simulations revealed that the jet becomes thinner and more intense as the Prandtl number increases. In particular, at moderate Reynolds numbers ($Re=UR/\nu$), the characteristic momentum and density radii of the jet scale as $(Fr/Re)^{1/2}$ and $(Fr/PrRe)^{1/2}$, respectively, while its length increases with either $Fr$ or $Pr$ \citep{hanazaki2015numerical,okino2023schmidt}. Regardless of the exact value of $Re$, increasing $Pr$ or lowering $Fr$ leads to a thinner jet and a stronger maximum upward velocity along the jet axis \citep{hanazaki2015numerical,zhang2019core}.

The collapse of the vortex ring and the emergence of an ascending jet in strongly stratified environments have notable consequences. In particular, a significant increase in the drag force experienced by the body is noticed. Early laboratory experiments \citep{abaid2004internal,camassa2009prolonged,yick2009enhanced} and resolved simulations \citep{torres2000flow,hanazaki2009schmidt,hanazaki2015numerical} attributed this drag enhancement to the distortion of the initially horizontal isopycnals caused by the descending sphere. This distortion leads to the downward entrainment of lighter fluid over a distance much larger than the body size, generating a substantial buoyancy force that resists the motion. However, various theoretical models based on the entrained volume of light fluid failed to predict quantitatively the correct magnitude of the drag enhancement observed under strong stratification conditions, \textit{i.e.}, $Fr \lesssim1$ \citep{srdic1999gravitational,candelier2014history,mehaddi2018inertial}, suggesting that additional mechanisms are involved. \citet{zhang2019core} made use of a rigorous decomposition of the velocity and pressure fields to quantify the individual contributions to the drag. Their analysis revealed that, within the range of parameters examined in both experiments and simulations, the dominant source of drag enhancement is not the additional buoyancy force resulting from fluid entrainment, but rather the specific structure of the vorticity field induced by buoyancy effects. These findings provide a comprehensive understanding of drag enhancement across a broad range of stratification conditions.

In contrast, what is much less well understood is the reason why, under highly stratified conditions ($Fr \lesssim 0.2$) and in the presence of weak enough viscous and diffusive effects, a three-dimensional (3D) instability emerges in the upward jet: the axisymmetric structure becomes unstable, yielding a coherent, meandering jet, as first observed by \citet{hanazaki2009jets}, and eventually leading to turbulence, as reported by \citet{akiyama2019unstable}. These authors suggested that the instability originates from shear-driven disturbances in the thin boundary layer surrounding the body, these disturbances being advected into the jet and subsequently amplified downstream. While these studies offer valuable qualitative insights, they do not provide a self-consistent explanation of the underlying instability mechanism. Experimentally, this gap in understanding arises from the difficulty in capturing the full 3D structure of the velocity and density fields. Computationally, resolving the extremely thin  boundary layer of the stratifying agent in the case of salt and the fine structure of the jet's core represents a major challenge. Specifically, the characteristic thickness of the density boundary layer scales as $\delta_\rho/R \sim Re^{-1/2}Pr^{-1/3}$ \citep{zhang2019core,okino2023schmidt}, while the characteristic jet radius scales as $\delta_j/R \sim (Fr/RePr)^{1/2}$ \citep{hanazaki2015numerical,okino2023schmidt}. For example, with $(Fr, Re, Pr) = (0.1, 200, 700)$, these estimates yield $\delta_\rho/R \sim 8 \times 10^{-3}$ and $\delta_j/R \sim 8 \times 10^{-4}$, indicating that spatial resolutions on the order of $\Delta/R \sim 10^{-3}$ to $10^{-4}$ are required to adequately capture the steep gradients in these regions, with $\Delta$ denoting the computational cell size. \\
\indent Achieving such resolutions remains a major challenge despite current computational capabilities. Even with advanced body-fitted and Cartesian grid techniques, resolving both the density boundary layer and the narrow jet necessitates significant computational resources and highly efficient parallel algorithms. As a result, the 3D jet instability has only been explored experimentally \citep{hanazaki2009jets, akiyama2019unstable} so far, while nearly all existing numerical studies have focused on two-dimensional (2D) axisymmetric configurations \citep{torres2000flow,hanazaki2009schmidt,hanazaki2015numerical,zhang2019core,okino2023schmidt}. Consequently, the mechanisms that trigger and sustain this instability, as well as its broader dynamical implications, remain poorly understood. However, the meandering of the jet obviously induces some lateral motion of the body if the latter is free to move, leading to non-vertical trajectories \citep{mercier2020settling}. Improving this still preliminary knowledge is the motivation of the present study. We aim to investigate the 3D jet instability in depth, by carrying out fully-resolved simulations over a wide range of the $(Fr, Re, Pr)$ parameters. Through this approach, we seek to uncover the physical mechanisms underlying the onset and saturation of the instability, and to provide a comprehensive description of how stratification influences the dynamics of the buoyant jet.\\
\indent To this aim, the present study employs the in-house JADIM code developed at IMFT to obtain fully-resolved 3D velocity and density fields. This code has been extensively used to investigate the dynamics of bubbles and rigid bodies in various flow configurations \citep{magnaudet2007wake, auguste2018path}. In particular, its robustness and accuracy were validated under conditions relevant to the present study by \citet{zhang2019core}, who examined the drag enhancement on a sphere settling in a linearly stratified fluid within a 2D axisymmetric framework over a broad range of parameters. Building on these foundations, results of the present 3D simulations enable a comprehensive analysis of the jet instability, including its spatio-temporal development and sensitivity to the $(Fr, Re, Pr)$ parameters. Most importantly, these results help us elucidate the physical mechanisms driving the instability. The rest of the paper is structured as follows. Section \ref{sec2} introduces the problem formulation and numerical methods. Section \ref{sec3} presents an overview of the numerical results, with a focus on the global evolution of the jet structure and the hydrodynamic forces on the body over a broad range of parameters. Section \ref{sec4} makes use of several diagnostics to provide a detailed analysis of the successive stages of the jet instability. Section \ref{sec5} investigates the physical mechanisms underlying the non-axisymmetric instability of the jet, and how these processes depend on the flow parameters. A summary of the main findings is provided in \S\,\ref{sec6}.

\section{Numerical approach}
\label{sec2}
\subsection{Problem statement and governing equations}
\begin{figure}
\vspace{5mm}
\centering
\includegraphics[width=0.9\textwidth]{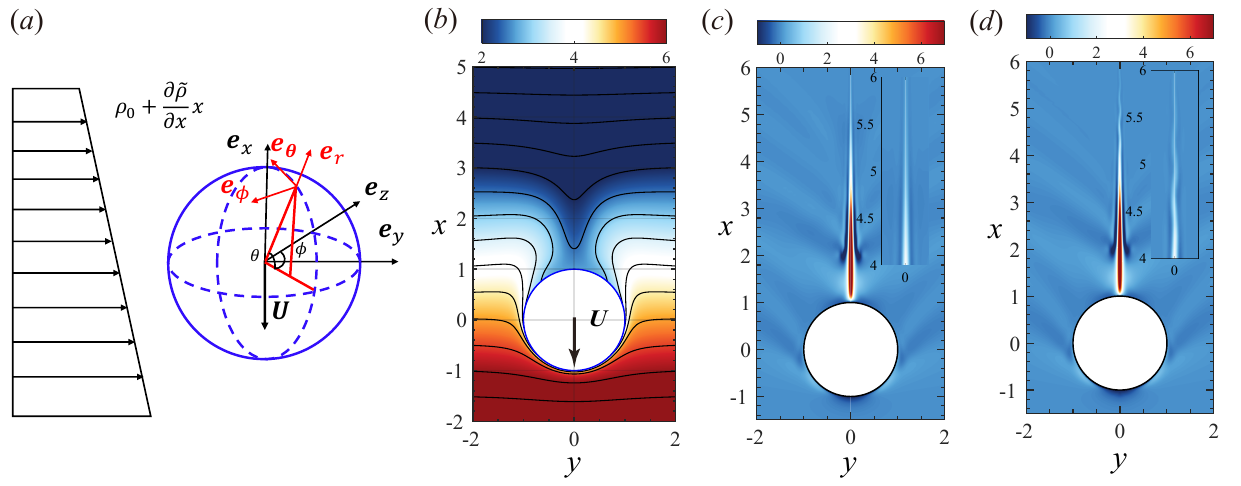}
\caption{$(a)$: Sketch of the flow configuration and definition of some quantities; $(b)$: light fluid dragged down by the sphere at $ (Fr,Re, Pr) = (1,100, 70)$, with the iso-contours and colors highlighting isopycnals and density variations; $(c)$: stable jet observed in a 2D axisymmetric simulation at $ (Fr,Re, Pr) = (0.1,100, 700)$; $(d)$: unstable jet observed with the same set of parameters in a 3D simulation. In $(c,d)$, colors refer to the absolute vertical velocity $\boldsymbol{u} \cdot \boldsymbol{e}_x - 1$; the inset provides an enlarged view of the upper part of the jet.}
\label{fig-i}
\end{figure}

We consider a rigid sphere with radius $R$ translating vertically with a constant velocity $-U\boldsymbol{e}_{x}$ through a linearly stratified fluid with a reference density $\rho_0$ and a vertical density gradient $\partial\tilde{\rho}/\partial x < 0$, with $\boldsymbol{e}_{x}$ denoting the upward unit vector (see figure~\ref{fig-i}$(a)$), and $\tilde{\rho}(x)$ representing the undisturbed background density profile. The fluid has uniform dynamic viscosity $\mu$ and molecular diffusivity $\kappa$. We normalize lengths, velocities, times, pressure and density variations using characteristic scales $R$, $U$, $R/U$, $\rho_0 U^2$, and $-R\,\partial\tilde{\rho}/\partial x$, respectively. In the body-fixed reference frame, the fluid velocity is denoted by $\boldsymbol{u}(\boldsymbol{x}, t)$ at position $\boldsymbol{x} = (x, y, z)$ and time $t$. The local fluid density is expressed as
\begin{equation}
\tilde{\rho}(x) + \rho(\boldsymbol{x},t)= \rho_{0} + t - x + \rho(\boldsymbol{x},t)\,,
\label{densi}
\end{equation}
where $\rho(\boldsymbol{x}, t)$ denotes the density disturbance resulting from the body motion.

We frequently use spherical coordinates to describe the flow field, with the polar and azimuthal directions, $\theta$ and $\phi$, as defined in figure~\ref{fig-i}$(a)$. Under the Boussinesq approximation, the non-dimensional governing equations for the velocity $\boldsymbol{u}$ and density disturbance $\rho$ take the form \citep{torres2000flow,hanazaki2015numerical}

\begin{eqnarray} \label{equ_2-1}
\nabla  \cdot \boldsymbol{u} = 0\,,
\end{eqnarray}
\begin{eqnarray} \label{equ_2-2}
\partial_t\boldsymbol{u} + \boldsymbol{u}\cdot\nabla\boldsymbol{u} =  - \nabla p - Fr^{-2}\rho{{\boldsymbol{e}}_x} + Re^{-1}{\nabla ^2}{\boldsymbol{u}}\,,
\end{eqnarray}
\begin{eqnarray} \label{equ_2-3}
\partial_t\rho + \boldsymbol{u}\cdot\nabla\rho= \boldsymbol{u}\cdot\boldsymbol{e}_{x} - 1 + Pe^{-1}{\nabla ^2}\rho\,.
\end{eqnarray}

Equations (\ref{equ_2-2}) and (\ref{equ_2-3}) involve the Reynolds and P\' eclet numbers, characterizing the relative importance of advective and diffusive effects in the momentum and density balances, respectively. These two numbers are defined as $Re = UR/\nu$ and $Pe = UR/\kappa$, where $\nu = \mu/\rho_0$ is the kinematic viscosity. Their ratio yields the Prandtl number, $Pr = \nu/\kappa$. Additionally, the ratio of inertial to buoyancy forces is quantified by the Froude number, $Fr = U/(N R)$, with $N = [-(g/\rho_0)\partial_x\tilde{\rho}]^{1/2}$ being the Brunt-V\"ais\"al\"a frequency and $g$ denoting  gravity. Several studies \citep{hanazaki2015numerical,okino2017velocity,akiyama2019unstable} defined the Reynolds number based on the sphere diameter $2R$, whereas the characteristic radius $R$ is selected here to maintain consistency with our previous work \citep{zhang2019core} and align with the definition of $Fr$. The hydrostatic pressure component, $-(gR/U^2)x + Fr^{-2}(x^2/2 - x t)$, is incorporated into the modified pressure $p(\boldsymbol{x}, t)$. At the sphere surface $\mathcal{S}$, the no-flux and no-slip boundary conditions hold, implying

\begin{flalign} \label{equ_2-4}
\boldsymbol{n}\cdot\nabla\rho = \boldsymbol{n}\cdot\boldsymbol{e}_{x} \text{\quad and \quad} \boldsymbol{u}=0 \text{\quad on $(\mathcal{S})$}\,,
\end{flalign}

where $\boldsymbol{n}$ stands for the unit normal directed into the fluid. As disturbances vanish in the far field, $\rho$ and $\boldsymbol{u}$ also obey

\begin{flalign} \label{equ_2-5}
\rho \xrightarrow{} 0 \text{\quad and\quad} \boldsymbol{u} \xrightarrow{} \boldsymbol{e}_{x} \text{\quad for\quad} ||\boldsymbol{x}||\xrightarrow{}\infty\,.
\end{flalign}

We are also interested in examining how the forces acting on the sphere vary with the governing parameters. Since the sphere is constrained to move in the $\bm{e}_x$-direction, the drag force is always parallel to $\bm{e}_x$, while a transverse (or lift) force due to the flow instability may arise in the horizontal $(\bm{e}_y,\bm{e}_z)$ plane. Therefore, the force components are defined as 
\begin{eqnarray}\label{e2.4}
   F_D = \bm{e}_x\cdot\int_\mathcal{S} \mathbb{T} \cdot \bm{n}~d\mathcal{S}\,, \quad
   F_{L,y} = \bm{e}_{y}\cdot\int_\mathcal{S} \mathbb{T} \cdot \bm{n}~d\mathcal{S}\,,\quad F_{L,z} = \bm{e}_{z}\cdot\int_\mathcal{S} \mathbb{T} \cdot \bm{n}~d\mathcal{S}\,, 
\end{eqnarray}
where $\mathbb{T} = -p\mathbb{I} + Re^{-1}(\nabla \boldsymbol{u} + \nabla \boldsymbol{u}^{\mathrm{T}})$ is the stress tensor, $\mathbb{I}$ denoting the unit tensor. Normalizing these force components by $\frac{\pi}{2} R^2 \rho_0 U^2 $ yields the drag and lift coefficients $C_D$ and $C_{L,y}$ (or $C_{L,z}$), respectively.
As the sphere settles through the stratified fluid, the isopycnals are deflected around it, as illustrated in figure~\ref{fig-i}$(b)$, which leads to an increase in the drag coefficient $C_D$. In the absence of flow instability, a steady-state density field is eventually established, in which advection and diffusion are in balance. When the stratification is sufficiently strong, a buoyant jet forms downstream of the sphere, as shown in the example of figure~\ref{fig-i}$(c)$. In this figure, resulting from a 2D axisymmetric simulation, contours of the absolute vertical velocity, $\boldsymbol{u} \cdot \boldsymbol{e}_x - 1$, reveal that the fluid moves significantly faster than the background flow in the wake region. Figure~\ref{fig-i}$(d)$ depicts a qualitatively different scenario obtained from fully 3D simulations with the same parameter set. The tail region of the jet is now unstable and meanders weakly, as emphasized in the enlarged view. The present study thus aims to characterize the onset, evolution, and parametric sensitivity of this 3D jet instability, and clarify the underlying physical mechanisms at stake.

\subsection{Numerical methods and grid design}
\label{2.2}
The JADIM code employs a finite-volume discretization of the governing equations (\ref{equ_2-1} - \ref{equ_2-3}) on a staggered grid, where velocity components are defined at cell faces, while pressure and density are stored at cell centers \citep{magnaudet1995accelerated}. Time integration combines a third-order explicit Runge-Kutta scheme for advective and source terms with a semi-implicit Crank-Nicolson method for diffusive terms, ensuring second-order accuracy in both time and space \citep{calmet1997large}. Incompressibility is enforced at each time step via a projection method: an intermediate velocity field is corrected by solving a Poisson equation for the pressure increment, ensuring that the final velocity field is divergence-free to machine precision. At high P\'{e}clet number, \textit{i.e.}, for large $Pr$, the central differencing of the advective density flux can generate nonphysical oscillations. To avoid such oscillations and preserve the monotonicity of the density disturbance, the advective flux $\nabla \cdot (\rho\boldsymbol{u})$ is discretized using a total-variation-diminishing (TVD) ``monotonized-central" scheme based on the van Leer limiter \citep{van1977towards}. 
The JADIM solver summarized above was extensively validated by \cite{zhang2019core}. In particular, the corresponding 2D axisymmetric simulations past a settling sphere were shown to accurately capture both the detailed structure of the thin jet that emerges at low Froude number and the associated drag enhancement. 

In the present study, the same numerical strategy is extended to the 3D framework. Computations are performed in a spherical coordinate system with a non-uniform grid distribution in both the radial ($r$) and polar ($\theta$) directions. Near the sphere surface, the minimum grid spacing in the radial direction, $\Delta_r$, and that in the polar direction near the upper pole, $\Delta_\theta$, are set to $\Delta_r = \Delta_\theta = 5\times 10^{-4}$. For the most demanding case considered, $(Fr,Re, Pr) = (0.02,150, 700)$, the density boundary layer is significantly thinner than the momentum one, with a characteristic thickness $\delta_\rho \sim Re^{-1/2}Pr^{-1/3}=\mathcal{O}(10^{-2})$, while the radius of the jet scales as $\delta_j \sim (Fr/RePr)^{1/2} =\mathcal{O}(10^{-3})$ \citep{hanazaki2015numerical,okino2023schmidt}. Therefore, the above resolution ensures that both the density boundary layer and the  inner jet structure are properly resolved. A grid independence study presented in appendix\,\ref{appA} confirms that doubling $\Delta_r$ and $\Delta_\theta$ does not alter the results. Influence of the azimuthal resolution is also considered in that appendix. It is established that selecting a uniform distribution with $\Delta_\phi=\pi/32$ ensures grid convergence. The spherical computational domain has a outer radius $r_{\text{max}} = 40$, a size that was shown by \cite{zhang2019core} to be sufficiently large to avoid artificial confinement effects, especially in highly stratified cases (very low $Fr$) where the jet is very short. The total number of cells in the domain is $(N_r, N_\theta, N_\phi) = (420, 200, 64)$, with the $r$- and $\theta$- directions consistent with the previously validated 2D axisymmetric setup. 
The overall grid arrangement is illustrated in figure~\ref{fig6-2}$(a)$. 

The Dirichlet boundary condition \eqref{equ_2-5} is imposed to the velocity field over the part of outer boundary $r = r_{\text{max}}$ extending from the lower pole to the edge of the wake region. Within this wake region (defined arbitrarily as the cone originating from the sphere centre and making a $60^\circ$ semi-angle with the upper part of the $\boldsymbol{e}_{x}$-axis), the non-reflecting outlet condition proposed by \citet{magnaudet1995accelerated} is imposed. To avoid spurious reflections of internal waves from the outer boundary, a linear Rayleigh damping strategy is applied to density disturbances within a sponge layer extending over the last five cells adjacent to the outer boundary in the $r$-direction \citep{slinn1998model,chongsiripinyo2017vortex}. Specifically, a damping term $-\psi(r)\rho$ is added to the right-hand side of \eqref{equ_2-3}, where the weighting function $\psi(r)$ increases quadratically from zero at the inner edge of the sponge layer to unity at $r = r_{\text{max}}$. This method was already shown to be effective by \cite{zhang2019core}.

Some numerical aspects specific to 3D simulations deserve to be mentioned. In the 2D axisymmetric study of \cite{zhang2019core}, the time history of the drag coefficient was monitored, and the flow field was considered converged when $C_D$-variations dropped below $0.2\%$ over the final 2000 time steps. In contrast, 3D simulations are inherently more complex due to the possible emergence of the jet instability, which introduces temporal and spatial fluctuations. To address this difficulty, we established distinct convergence criteria based on the flow stability. For stable flows, \textit{e.g.}, at $Fr \geq 0.3$ for $(Re, Pr) = (100, 700)$, the same $C_D$-based strategy is still applied. Conversely, for strongly unstable flows, \textit{e.g.}, at $Fr \leq 0.05$, we rely on a criterion based on the lift coefficient, $C_L$. The flow is considered statistically stationary once $C_L$ exhibits periodic saturation over the last ten oscillation cycles (see panels $(c)$ and $(f)$ in figure~\ref{fig4-1}). In contrast, in intermediate regimes, \textit{e.g.}, $0.05 < Fr < 0.3$, the instability may evolve very slowly, and the flow does not reach convergence even beyond $t > 80$. These cases are therefore excluded from the quantitative analysis presented later. Moreover, we define the onset of jet instability based on a threshold on the azimuthal velocity component: the jet is said to be unstable when $|u_\phi^\mathrm{max}| > 10^{-2}$ close to its centreline. Cases with lower values of $|u_\phi^\mathrm{max}|$ are considered stable. 

In most simulations, the jet instability arises naturally due to the amplification of ambient numerical disturbances. Nevertheless, in regimes close to the onset of instability,  \textit{e.g.}, $Fr \approx 0.1$ for $(Re, Pr) = (100, 700)$, the natural development of disturbances is so slow that it is desirable to accelerate the growth of the instability with the help of an external disturbance. Similarly, in the kinetic energy budget analysis presented in \S\,\ref{sec4.2}, artificial disturbances are required to properly estimate the growth rate. In these cases, the disturbance is introduced only during the first 5000-20000 time steps beyond the time required to reach saturation in the equivalent 2D axisymmetric case (depending on the time step, it is applied over a time period $ \tau \sim 1-2$); then the computation proceeds naturally. The disturbance is imposed on the vertical velocity field $\boldsymbol{u}\cdot\boldsymbol{e}_{x}$ in the form $u_x' = u_0' \exp\left\{-[(y - y_0)^2 + (x - x_0)^2] / \epsilon^2\right\}$, with amplitude $u_0' = 10^{-4}$, central location $x_0 = 2, y_0 = 0.005$, and decay width $\epsilon = 0.05$. 
Appendix~\ref{appB} establishes that this artificial disturbance does not alter the intrinsic nature of the flow, even when it is given a much larger amplitude, \textit{i.e.} $u_0' = 10^{-2}$: the jet initially exhibits transient oscillations but subsequently relaxes back to its initial axisymmetric structure after the disturbance is removed.\\
\indent Last, we employed a post-processing technique based on the tracking of trajectories of massless Lagrangian particles to investigate the flow behaviour in the sphere's wake. Each particle evolves according to the kinematic relation $\mathrm{d}_t\mathbf{x}_p=\mathbf{u}(\mathbf{x}_p,t)$ where $\mathbf{x}_p$ denotes the instantaneous particle position. The velocity $\mathbf{u}(\mathbf{x}_p, t)$ at the particle location is obtained by interpolating the flow field over the surrounding $2 \times 2 \times 2$ grid cells. Then, starting from the particle position at time $n\Delta t$ (with $\Delta t$ denoting the time step), the position at time $(n+1)\Delta t$ is obtained explicitly as $\mathbf{x}_p^{\,n+1}=\mathbf{x}_p^{\,n}+\mathbf{u}(\mathbf{x}_p,t)\,\Delta t$. 

\section{Overview of numerical results}
\label{sec3}

\begin{figure}
\vspace{5mm}
\centering
\includegraphics[width=0.9\textwidth]{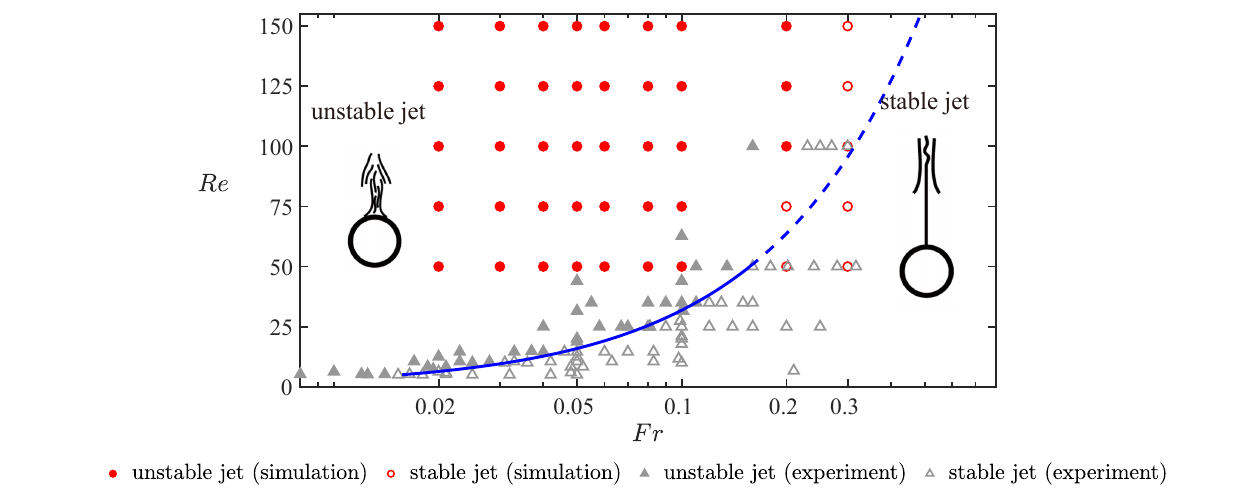}
\caption{Parameter range $(Re, Fr)$ investigated in simulations, compared with experiments by \citet{akiyama2019unstable} at $Pr=700$. Triangles and circles correspond to experimental and numerical observations, respectively. Open and closed symbols refer to stable and unstable jets, respectively. The solid blue line is the experimentally determined stability threshold in the range $5\leq Re \leq 50$, given, with present normalisation, by $Fr/Re \approx 3.14 \times 10^{-3}$.}
\label{fig4-0}
\end{figure}

Numerical studies have revealed several key transitions in the wake of a sphere descending vertically through a stratified fluid. In particular, 2D axisymmetric simulations by \citet{torres2000flow,hanazaki2009schmidt,hanazaki2015numerical,zhang2019core} showed that the standing eddy that takes place at the back of the sphere when $Re\gtrsim10$ in an unstratified fluid ($Fr=\infty$) collapses into a thin vertical jet as the Froude number decreases below a critical value. This threshold strongly depends on the Reynolds and Prandtl numbers. Experimental investigations \citep{hanazaki2009jets, akiyama2019unstable} confirmed that the axisymmetric jet becomes unstable and starts meandering when $Fr$ further decreases to a second, lower threshold. 

\citet{akiyama2019unstable} summarized their observations in a $Re-Fr$ phase diagram at a fixed Prandtl number $Pr = 700$, corresponding to salinity-induced stratification in water. Their observations, covering the range $0.003 \leq Fr \leq 1$ and $2 \leq Re \leq 50$, are reproduced in figure~\ref{fig4-0}. 
Present numerical observations spanning the range $0.02 \leq Fr \leq 0.3$ and $50 \leq Re \leq 150$ are plotted in the same figure. Lower Reynolds numbers, as explored in experiments, were not investigated here because the onset of jet instability is already clearly observed at $Re = \mathcal{O}(100)$, while capturing the instability at lower Reynolds numbers ($Re = \mathcal{O}(10)$) would require significantly longer computational times. Both the experimental and numerical data plotted in figure~\ref{fig4-0} indicate that the stability criterion $Fr/Re \approx 3.14 \times 10^{-3}$ proposed by \citet{akiyama2019unstable} (dashed blue line) becomes invalid beyond $Re > 50$ or $Fr>0.15$. In particular, both datasets identify that the transition at $Re = 100$ takes place in the range $0.2<Fr <0.3$, while this criterion predicts a critical value of $0.314$. Understanding the unstable regime highlighted in figure~\ref{fig4-0} has implications beyond the wake dynamics of settling spheres. For example, \citet{mercier2020settling} demonstrated that the meandering jet behind a freely falling disk can trigger path instability, significantly affecting the settling dynamics. These findings emphasize the need to better understand how the jet structure evolves when the flow parameters are varied. \\
\indent We also carried out numerical simulations with Prandtl numbers $Pr = 0.7,\ 7$ and $ 70$, to cover especially the diffusion of heat in air and water under standard conditions. No instability was observed at $Pr = 0.7$, due to the strong diffusive effects. This is why the following discussion focuses on $Pr = 7,\ 70$ and (mostly) $700$.

\subsection{Evolution of the transverse force stemming from the unstable jet}\label{sec3.1}

\begin{figure}
\vspace{5mm}
\centering
\includegraphics[width=0.9\textwidth]{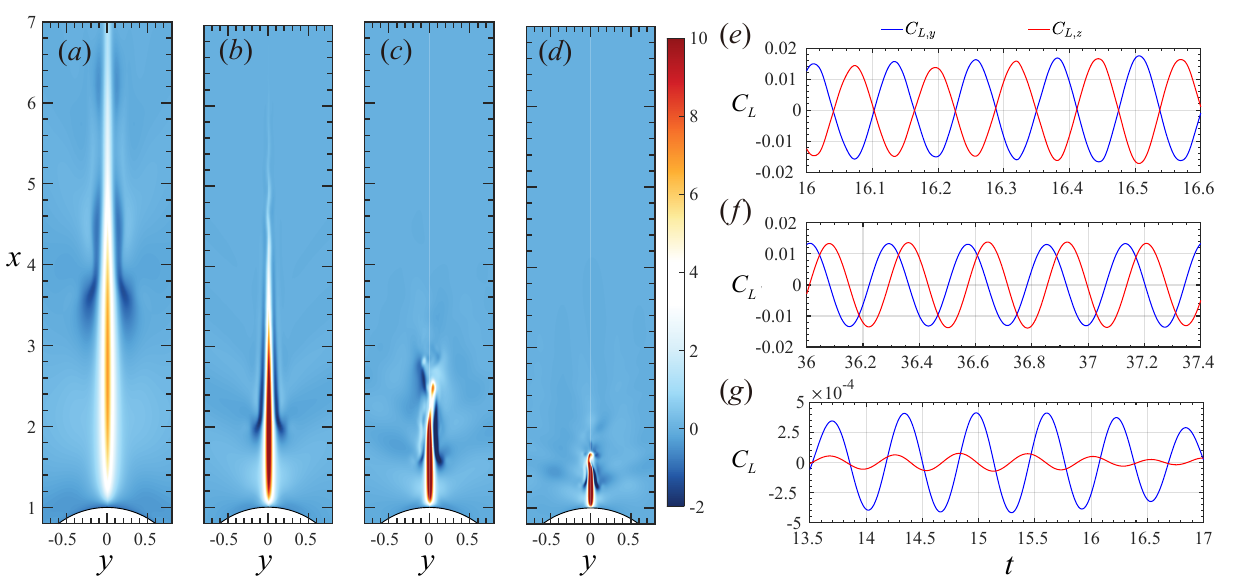}
\caption{Examples of the jet evolution and instability characteristics observed by varying $Fr$, with $Re=100$ and $Pr=700$ in all cases. $(a-d)$: contours of the vertical velocity projected onto the vertical cross-sectional $(x,y)$ plane, with $(a)$: $Fr = 0.3$ (stable); $(b)$ $Fr = 0.1$ (chaotic instability); $(c)$: $Fr = 0.05$ (spiral instability); $(d)$: $Fr = 0.02$ (standing wave instability). $(e-g)$: time histories of the $y$- and $z$-components of $C_L$ during the saturated oscillation stage, with $(e)$: $Fr = 0.1$; $(f)$: $Fr = 0.05$; $(g)$: $Fr = 0.02$.}
\label{fig4-1}
\end{figure}

We first fix the Reynolds and Prandtl numbers at $(Re, Pr) = (100, 700)$ and progressively reduce the Froude number, starting from the critical curve shown in figure~\ref{fig4-0}. Figures~\ref{fig4-1}$(a-d)$ present colormaps of the absolute streamwise velocity $u_x = \bm{u}\cdot\bm{e}_x-1$ projected onto the vertical cross-sectional plane $(x,y)$ for $Fr = 0.3$, $0.1$, $0.05$, and $0.02$, respectively. In all cases, a high-speed jet forms in the wake, consistent with previous numerical findings \citep{torres2000flow, hanazaki2015numerical, zhang2019core} and experimental observations \citep{hanazaki2009jets}. A bell-shaped structure (dark blue region) is also visible along the jet axis, as previously described by \citet{hanazaki2015numerical}. These authors showed that this peculiar structure is associated with internal waves generated in the wake. As stratification intensifies, \textit{i.e.}, $Fr$ decreases, the jet becomes progressively shorter and thinner, in agreement with earlier observations. 

As shown in figure~\ref{fig4-1}$(a-d)$, while the jet and bell-shaped structure remain axisymmetric at $Fr = 0.3$, oscillations emerge for $Fr \leq 0.1$, indicating the onset of the 3D instability, consistent with experimental observations \citep{hanazaki2009jets, akiyama2019unstable}. At $Fr = 0.1$, a weak meandering is visible near the jet tip at the vertical position $x \approx 4.5$, while the bell-shaped structure remains axisymmetric. As stratification increases further, i.e., $Fr = 0.05$ and $0.02$, the jet exhibits stronger oscillations that take place closer to the sphere, at $x \approx 2.5$ and $1.6$, respectively. The meandering behaviour may be characterized by the lift coefficient, $C_L$. Figures~\ref{fig4-1}$(e-g)$ show the time histories of the $y$- and $z$-components of $C_L$ during the saturated oscillatory regime for the three unstable cases. These plots reveal sinusoidal oscillations in both $C_{L,y}$ and $C_{L,z}$, with amplitudes increasing from $\mathcal{O}(10^{-4})$ at $Fr = 0.1$ to $\mathcal{O}(10^{-2})$ at $Fr = 0.05$ and $0.02$. This increase likely arises from two factors, namely the shorter distance between the meandering jet and the sphere and the increase in the oscillation amplitude as $Fr$ decreases.

\begin{figure}
\vspace{5mm}
\centering
\includegraphics[width=0.96\textwidth]{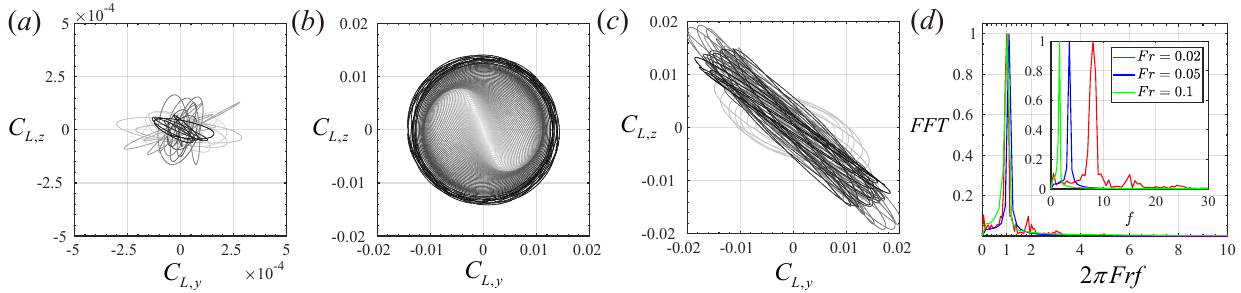}
\caption{Wake evolution tracked in the $(C_{L,y},C_{L,z})$ phase plane in three distinct unstable regimes, with $Re=100$ and $Pr=700$ in all cases. $(a)$: $Fr = 0.1$ (chaotic regime); $(b)$: $Fr = 0.05$ (spiral regime); $(c)$: $Fr = 0.02$ (standing wave regime); $(d)$: Fourier transforms (FFT) of the lift coefficient $C_{L,y}$ for different stratification levels. In $(a-c)$, the color along the path transitions from grey to black following time progression. }
\label{fig4-2}
\end{figure}

To better quantify the jet instability, we examine the time evolution of the lift coefficients in the $(C_{L, y},C_{L, z})$ phase space, as shown in figure~\ref{fig4-2}$(a-c)$. At $Fr = 0.1$, the path exhibits chaotic excursions throughout the phase plane. Although the lift coefficient remains small ($\mathcal{O}(10^{-4})$), this behaviour is not due to numerical noise, since simulations with weaker stratifications (higher $Fr$) indicate that $C_L$-variations remain of $\mathcal{O}(10^{-6})$. Similar chaotic dynamics have been reported in unstratified flows past bluff bodies, such as circular disks \citep{shenoy2008flow,auguste2009bifurcations} and rigid spheroids \citep{chrust2010parametric}. Such transitions are often associated with irregular vortex shedding patterns. We did not examine this possibility in detail, nor did we attempt to confirm more rigorously the chaotic nature of the observed motion. \\
\indent At $Fr = 0.05$, the $(C_{L, y},C_{L, z})$ trajectory exhibits a spiral structure, preceded by planar oscillations. We refer to this behaviour as the ``spiral" mode. Interestingly, similar spiral paths of transverse force coefficients have been reported in counter-flows past heated spheres \citep{kotouvc2009transition}, where local buoyancy forces induce a similar symmetry-breaking behaviour. In that case, two longitudinal vortex threads were found to rotate slowly around each other, corresponding to a gradual rotation of the wake's symmetry plane. Notably, the peak amplitude observed here in the lift coefficients ($C_L^{\mathrm{max}} \approx 0.014$) is comparable to that reported by the previous authors,  namely $C_L^{\mathrm{max}} = 0.019$ (see their figure 13). As the stratification becomes stronger ($Fr = 0.02$), the $(C_{L, y},C_{L, z})$ oscillations transition to a periodic regime with zero-mean value, confined to a fixed symmetry plane. We refer to this regime as the ``standing wave" mode. Similar patterns have been observed in unstratified flows past thin disks and spheres \citep{shenoy2008flow, chrust2010parametric}, and are distinct from the classical planar ``zig-zag" mode, which also preserves planar symmetry but exhibits a non-zero mean lift force \citep{fabre2008bifurcations, meliga2009global}. Nevertheless, the peak amplitude in the lift coefficients is significantly smaller in the present case, with $C_L^{\mathrm{max}} \approx 0.02$, to be compared with $C_L^{\mathrm{max}} = 0.042$ in the uniform flow past a circular cylinder at $Re = 90$ \citep{shenoy2008flow} or $C_L^{\mathrm{max}} = 0.069$ for a sphere at $Re = 150$ \citep{johnson1999flow}. Figure~\ref{fig4-2}$(d)$ presents the Fourier spectra of the lift coefficients for different levels of stratification. In all cases, the dominant peak closely matches the Brunt-V\"ais\"al\"a frequency, such that the normalized oscillation frequency $2\pi Fr(fR/U)$ is close to unity. This correspondence suggests that the fluctuations in the transverse force, hence the jet meandering, are intimately linked to the internal waves generated by the body as it descends through the fluid.

\begin{figure}
\vspace{5mm}
\centering
\includegraphics[width=0.96\textwidth]{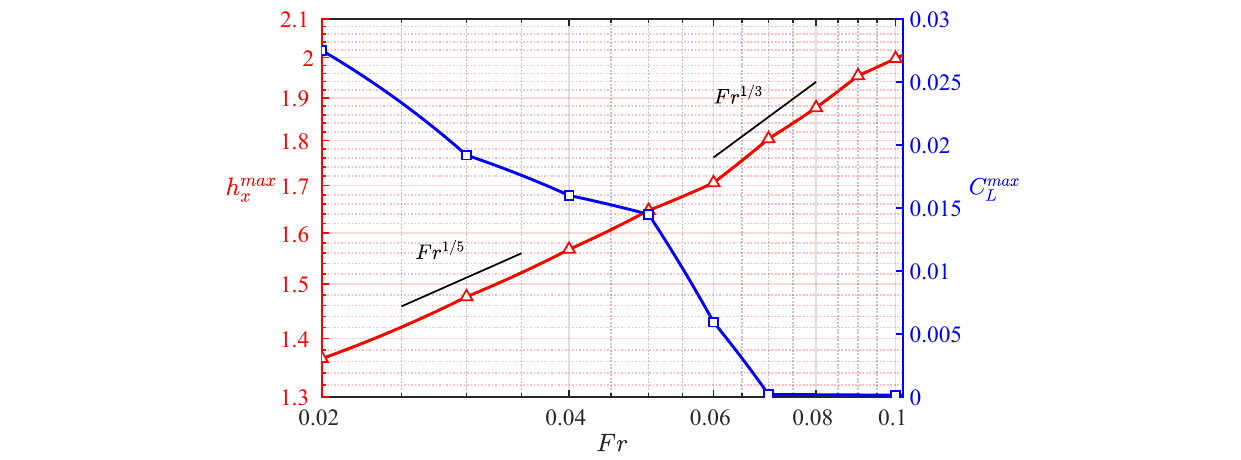}
\caption{Variation with the Froude number of the maximum amplitude of the transverse force, $C_L^{\mathrm{max}}$ (blue line, right axis), and the axial location $h_x^{\mathrm{max}}$ of the peak vertical velocity on the jet axis in the axisymmetric configuration (red line, left axis). 
}
\label{fig4-5}
\end{figure}

Additional simulations for stratification levels in between $Fr = 0.05$ and $Fr = 0.02$ reveal that the $(C_{L, y},C_{L, z})$ trajectory exhibits a flattened spiral shape at $Fr = 0.03$ (not shown). The successive transitions observed from $Fr=0.1$ to $Fr=0.02$ indicate that, as stratification intensifies, the meandering jet becomes increasingly confined to a 2D vertical plane. Following the interpretation of \citet{kotouvc2009transition} in the context of heated spheres, the emergence of the spiral mode is attributed to an instability in the far-wake region, where the vorticity threads are less constrained and allowed to twist due to a slow rotation of the wake's symmetry plane. This motivates an investigation of the relationship between the observed instability ``style" and the axial location $h_x^{\mathrm{max}}$ at which the velocity reaches its maximum on the jet axis in the corresponding axisymmetric configuration. Figure~\ref{fig4-5} quantifies this connection across a range of Froude numbers. 
As $Fr$ decreases,  $h_x^{\mathrm{max}}$ also decreases. The plot suggests that this decrease follows a scaling transition from $Fr^{1/5}$ to $Fr^{1/3}$ as $Fr$ increases, but present results only cover one decade in $Fr$ so that this prediction must be considered with caution. The trend in figure~\ref{fig4-5} coincides with a transition in the instability behaviour - from a spiral mode originating in the far wake to a standing wave mode  initiated closer to the sphere - suggesting that the spatial location at which the jet destabilizes plays a critical role in shaping the ensuing dynamics. Meanwhile, the maximum amplitude of the transverse force, $C_L^{\mathrm{max}}$, decreases monotonically with $Fr$, reinforcing the view that stronger stratification leads to earlier jet destabilization and greater intensity of jet oscillations.
\subsection{Wake structure}\label{sec3.2}

\begin{figure}
\vspace{4mm}
\centering
\includegraphics[width=0.96\textwidth]{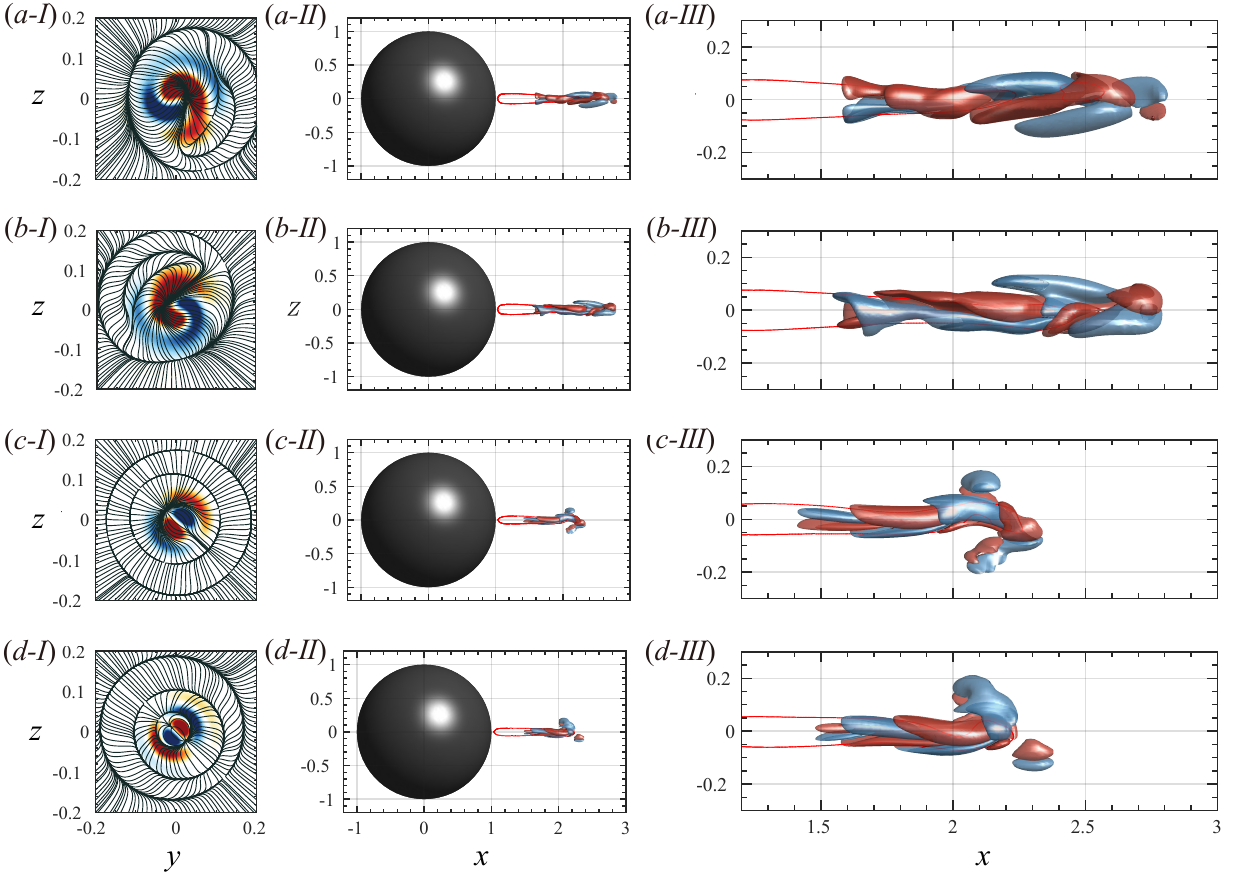}
\caption{Vortical structure in the sphere's wake. $(a-b)$: spiral mode at $Fr = 0.05$ shown at two successive time instants, $t = 30$ and $t = 30.1$; $(c-d)$: standing wave mode at $Fr = 0.02$ shown at $t = 17.09$ and $t = 17.11$. In each row, panel $(I)$ presents contours of the axial vorticity $\omega_x$ in a colour scale ranging from dark blue ($\omega_x=-3$) to dark red ($\omega_x=+3$), and streamlines in the cross-sectional plane $x = 1.8$ in $(a-b)$ and $x = 1.6$ in $(c-d)$; panel $(II)$ shows the 3D vortical structure visualized by iso-surfaces $\omega_x = \pm 3$ in $(a-b)$ and $\omega_x = \pm 20$ in $(c-d)$, with the vertical velocity iso-contour $(u_x - 1) = 1.2$ highlighted in red; panel $(III)$ is a close-up view of the wake structure shown in $(II)$.}
\label{fig4-3}
\end{figure}

Figure~\ref{fig4-3} presents a typical example of the instantaneous wake structure in the spiral and standing wave modes. 
Visualizations in panels $(a-b)$ reveal the helical structure of the spiral mode, with two longitudinal vortex threads wrapped around each other. 
A similar structure was identified by \citet{kotouvc2009transition} in the wake of a heated sphere. 
According to their interpretation, advective effects in the near wake act to stabilize the flow and suppress vortex shedding, whereas further downstream the wake becomes more prone to instability. Here, stratification effects at $Fr = 0.05$ are still moderate enough to make the jet meander some distance downstream of the sphere, which provides the spatial extent needed for the the symmetry plane of the wake to rotate slowly.

This is no longer the case at $Fr=0.02$ (panels $(c-d)$), where the two counter-rotating longitudinal vortices stay in a fixed symmetry plane. The vortex structure exhibits a clear periodic shedding, in agreement with the oscillatory behaviour of the transverse force observed in figure~\ref{fig4-2}$(c)$. The symmetry plane is seen to align with an azimuthal angle of $\theta = 45^\circ$ in the $(y,z)$ axes, consistent with the dominant direction of lift oscillations observed in figure~\ref{fig4-2}$(c)$. This further supports the classification of the observed instability as a standing wave mode. The wake structure closely resembles that found in some regimes in unstratified flows past spheres and disks \citep{shenoy2008flow, kotouvc2009transition, chrust2010parametric}. These findings establish the one-to-one connection between the various oscillatory modes observed in the transverse force and the characteristics of the vortex structures in the wake.
In the present study, the sphere is forced to translate vertically with a constant speed. If it were free to move under buoyancy effects, its path would be influenced by this unsteady wake dynamics. Therefore, it is expected that the distinct oscillatory wake modes identified above would translate into different styles of rise or fall of the body. 

\section{Successive stages of the jet instability}
\label{sec4}

\begin{figure}
\vspace{3mm}
\centering
\includegraphics[width=0.96\textwidth]{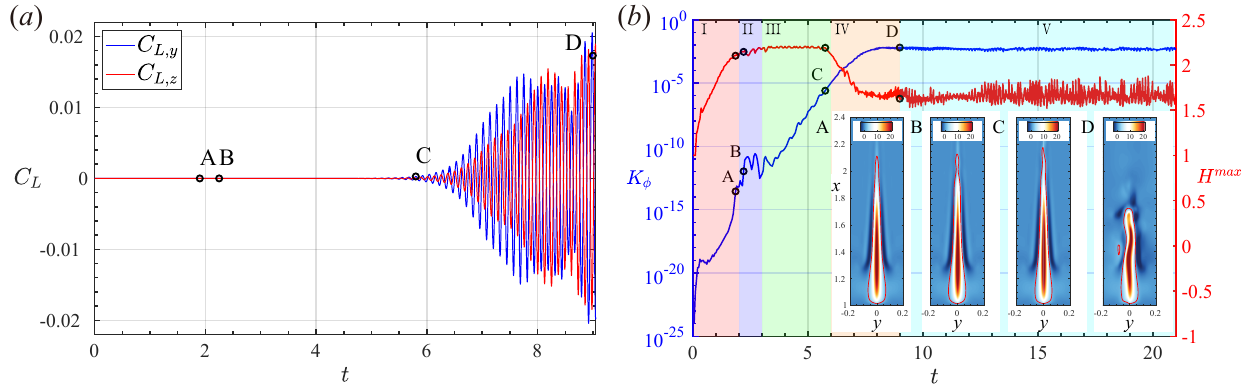}
\caption{Development of the jet instability for $(Fr,Re, Pr) = (0.02,100, 700)$, the reference case used throughout \S\S\,\ref{sec4} and \ref{sec5} unless specified otherwise. $(a)$: time evolution of the $y$- and $z$-components of the transverse force; $(b)$:  evolution of the azimuthal kinetic energy, $K_\phi$ (blue line, left axis), and the vertical position of maximum asymmetry, $H^\text{max}$ (red line, right axis). Five distinct stages of the flow in the wake region are identified, labelled I-V and identified by coloured regions in $(b)$. Insets show the jet structure in the vertical cross-sectional plane $(x,y)$ in the first four regimes, visualized by identifying the flow region where $(0 \leq |u_x - 1| < 22)$, at selected time instants $A, B, C$, and $D$.}
\label{fig5-1}
\end{figure}

In this section, we investigate in more detail how the jet instability arises. We focus on the parameter set $(Fr,Re, Pr) = (0.02,100, 700)$, and discuss other cases when relevant. Figure~\ref{fig5-1}$(a)$ shows the time evolution of the $y$- and $z$-components of the transverse force. Both components are seen to evolve from zero to an almost saturated value around $t=9$, revealing the development of the 3D instability. Figure~\ref{fig5-1}$(b)$ presents the evolution of two other quantities of interest to characterize the nature of the bifurcation taking place in the jet, namely the kinetic energy $K_\phi$ associated with the azimuthal velocity component (blue line), and the vertical position $H^{\text{max}}$ (red line) at which the wake asymmetry reaches its maximum. Specifically, the azimuthal kinetic energy is defined as 
\begin{equation}
K_\phi = \frac{1}{2} \int_\Omega  u_\phi^2 d\Omega\,, 
\label{eq_kphi}
\end{equation}
where $\Omega$ denotes the entire fluid domain. The quantity $H^\text{max}$ is computed by evaluating the standard deviation of the radial position $r_{1.2}(\phi,x)$ of the velocity iso-surface $(u_x - 1) = 1.2$ at a fixed vertical position $x$ over all azimuthal directions, namely
\begin{eqnarray}
\sigma_{x}(x) = \left[\frac{1}{2\pi} \int_{0}^{2\pi} r_{1.2}^2(\phi,x)d\phi\right]^{1/2}, \qquad H^\text{max} = \underset{x>1}{\arg\max}  \left( \sigma_{x}(x) \right)\,.
\label{eq_Hmax}
\end{eqnarray}

Based on the evolution of these metrics, five distinct stages of the flow are identified in figure~\ref{fig5-1}$(b)$:

\begin{itemize}
  \item \textit{\textbf{Stage $I$, 2D axisymmetric flow}} ($t < 2$). In this early stage, $K_\phi$ increases rapidly but remains very small ($K_\phi < 10^{-10}$), so that the flow remains essentially axisymmetric. This is confirmed by the inset in figure~\ref{fig5-1}$(b)$ at time A, where the jet structure is displayed. Meanwhile, $H^\text{max}$ rises monotonically, indicating that the most asymmetric region of the wake moves away from the sphere's upper pole.
  \item \textit{\textbf{Stage $II$, varicose instability}} ($2 < t < 3$). This stage is characterized by fluctuations in both $K_\phi$ and $H^\text{max}$. The inset at time B shows the jet starting to pinch off near its tail ($x \approx 2.0$). Nevertheless, the flow remains essentially axisymmetric, as the low values of $K_\phi$ (still below $10^{-10}$) confirm. As will be discussed later, the jet exhibits periodic stretching, necking, and pinch-off, consistent with the features of a convectively unstable flow. 
  \item \textit{\textbf{Stage $III$, onset of the sinuous instability}} ($3 < t < 6$). Now, $K_\phi$ grows nearly exponentially, increasing by seven orders of magnitude from the beginning to the end of this stage. 
  Meanwhile, $H^\text{max}$ reaches a plateau, implying that the instability is no longer convected downstream. Instead, it now grows at a fixed vertical position, a distinctive feature of absolute instabilities. The inset at time C reveals a clear meandering of the jet, consistent with experimental observations \citep{hanazaki2009jets, akiyama2019unstable}. We refer to the observed behaviour as the sinuous instability, aligning with similar behaviours in liquid jets and diffusion flames.
  \item \textit{\textbf{Stage $IV$, saturation of the sinuous instability}} ($6 < t < 9$). This stage is marked by a significant decrease in the growth rate of $K_\phi$, which eventually saturates around a mean value of $\mathcal{O}(10^{-2})$, owing to the increasing influence of nonlinear effects. 
  Concurrently, $H^\text{max}$ recedes from $\approx2.2$ to $\approx 1.7$, indicating that the region where the jet meanders most gets closer to the sphere. This is confirmed by the inset at time D, which shows that the sinuous instability has propagated upstream in between the two stages.
  \item \textit{\textbf{Stage $V$, saturated sinuous instability}} ($t > 9$). In this final stage, both $K_\phi$ and $H^\text{max}$ have already reached nearly constant values and only exhibit small-amplitude, high-frequency oscillations around these values. The jet continues to meander and its structure does not exhibit any significant difference with that observed at time D; this is why no additional inset is included in figure \ref{fig5-1}$(b)$.
\end{itemize}

Since regime $I$ was already thoroughly investigated in prior numerical works \citep{torres2000flow, hanazaki2009schmidt, hanazaki2015numerical, zhang2019core, okino2023schmidt}, the following sections focus on regimes $II$ to $V$ to provide a comprehensive picture of the development of the jet instability. 

\subsection{Stage $II$: Varicose instability}\label{sec4.1}

\begin{figure}
\vspace{3mm}
\centering
\includegraphics[width=0.96\textwidth]{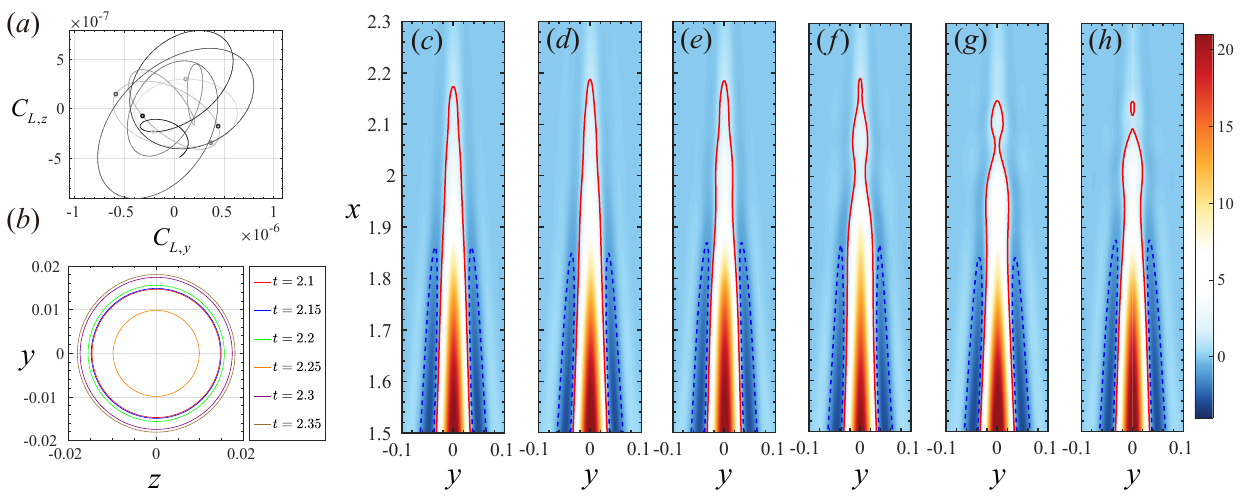}
\caption{Stage II: varicose instability. $(a)$: trajectory of the transverse force components in the $(C_{L,y},C_{L,z})$ phase plane for $2 < t < 3$, with circles indicating the time instants corresponding to the snapshots shown in $(b)$; $(b)$: horizontal cross sections of the iso-surface $(u_x - 1) = 1.2$ at $x = 2.1$ at selected times instants; $(c-h)$: distribution of the vertical velocity in the cross-sectional $(x,y)$ plane at the same instants of time. In $(c-h)$, the solid red and blue dashed lines denote the iso-contours $(u_x - 1) = 1.2$ and $(u_x - 1) = -1.4$, respectively.}
\label{fig5-3}
\end{figure}

Figure~\ref{fig5-3}$(a)$ shows that in the varicose stage, both components of the transverse force remain below $10^{-6}$ up to $t =2.6$, indicating a nearly axisymmetric wake. This is corroborated by the horizontal cuts of the iso-surfaces of the absolute vertical velocity displayed in panel $(b)$, which retain circular shapes undergoing an expanding-contracting-expanding sequence. The velocity contours in panels $(c-h)$ capture this process: the initial sword-like tip region (panels $(c-d)$) transitions to a tip region with a pronounced neck (panels $(e-f)$) before exhibiting clear signs of pinch-off (panels $(g-h)$). Such axisymmetric bulging-necking-bulging dynamics, typical of varicose modes, is widely documented for liquid jets \citep{sato1960stability,hussain1980controlled,huang1999mode} and flickering diffusion flames \citep{cetegen2000experiments,zhang2021instability,zhang2024experimental}. The behaviour observed here closely resembles that of such flames, in which the buoyancy-induced outer vortex ring plays a key role in triggering the varicose instability \citep{zhang2024experimental}.

\begin{figure}
\vspace{0mm}
\centering
\includegraphics[width=0.96\textwidth]{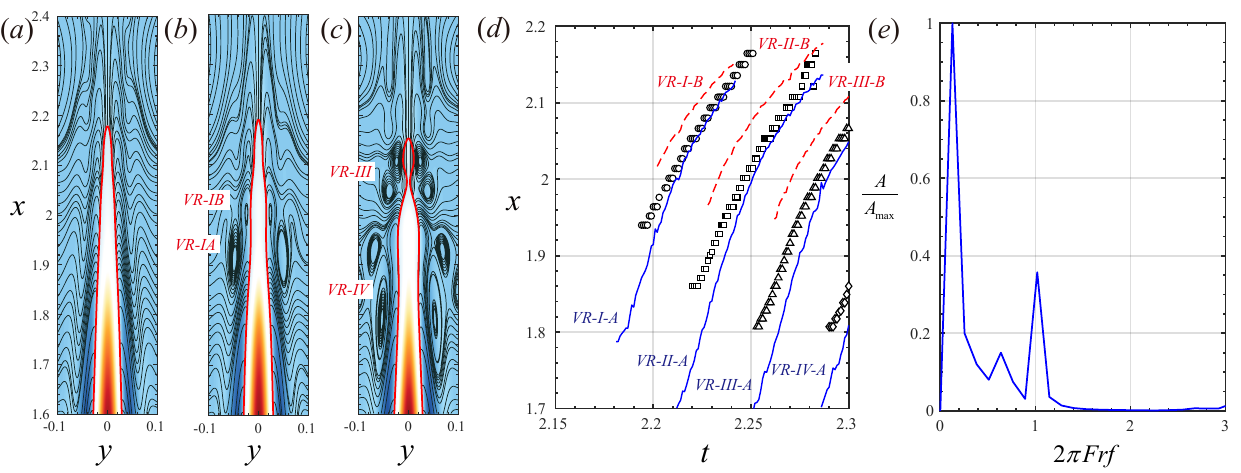}
\caption{Formation of vortex rings and interactions with the jet during the varicose stage (stage II). $(a-c)$: streamlines (in the laboratory frame) outside the jet at $t = 2.1$, $2.2$, and $2.3$, respectively; $(d)$: axial positions of the jet's neck (symbols) and vortex rings (lines) over time, with solid and dashed lines indicating the lower (A) and upper (B) ring in each pair; $(e)$ Fast Fourier transform of the drag coefficient during this stage. 
In $(b)$, VR-IA and VR-IB mark the first pair of vortex rings; in $(c)$, VR-III and VR-IV denote the third and fourth pairs of vortex rings, respectively, and the neck position is seen to lie in between the two rings of the VR-III pair.}
\label{fig5-4}
\end{figure}

Figures~\ref{fig5-4}$(a-c)$ show the emergence of vortex rings around the jet. At $t=2.1$, the velocity streamlines remain open. By $t=2.2$, two vortex rings form. The first of them, VR-IA, centered at $x = 1.92$, results from the pinching of the neighbouring streamlines, whereas VR-IB, centered at $x = 2.0$, is likely induced by the squeezing of fluid elements in between VR-IA and the jet. At $t=2.3$, the previous vortex pair and the next one (VR-II) have been advected downstream and are no longer visible in the spatial window displayed in panel $(c)$. In contrast, the next two pairs are, and the jet is seen to neck in between the two rings of the VR-III pair. The evolution of the axial position of the jet's neck and those of the centres of the successive vortex rings are displayed in figure~\ref{fig5-4}$(d)$, revealing that the former closely follows the latter. This in turn suggests that these vortex structures drive the necking and pinch-off of the jet. Figure~\ref{fig5-4}$(e)$ shows the fast Fourier transform of the drag force acting on the sphere. It reveals two dominant peaks, one at $f_1=1.67$ corresponding to the shedding frequency of the vortex rings, and a second one at $f_2=8.0$ matching the Brunt-V\"ais\"al\"a 
frequency (see figure \ref{fig4-2}$(d)$). The drag fluctuations induced by the shedding of vortex rings highlight the role of these specific flow structures in the stress distribution at the sphere surface. 

\begin{figure}
\vspace{3mm}
\centering
\includegraphics[width=0.96\textwidth]{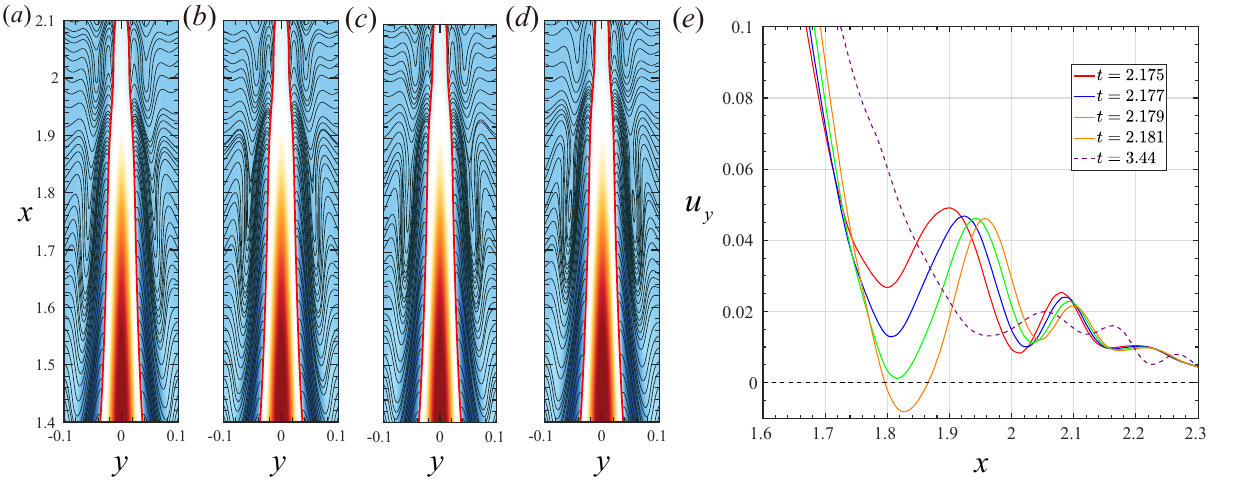}
\caption{Formation of vortex rings. $(a-d)$: evolution of the streamlines (in the laboratory frame) at $t = 2.175$, $2.177$, $2.179$, and $2.181$, respectively; $(e)$: evolution of the radial velocity component along the vertical line $y = 0.05, z=0$.} 
%
\label{fig5-6}
\end{figure}
\begin{figure}
\centering
\includegraphics[width=0.96\textwidth]{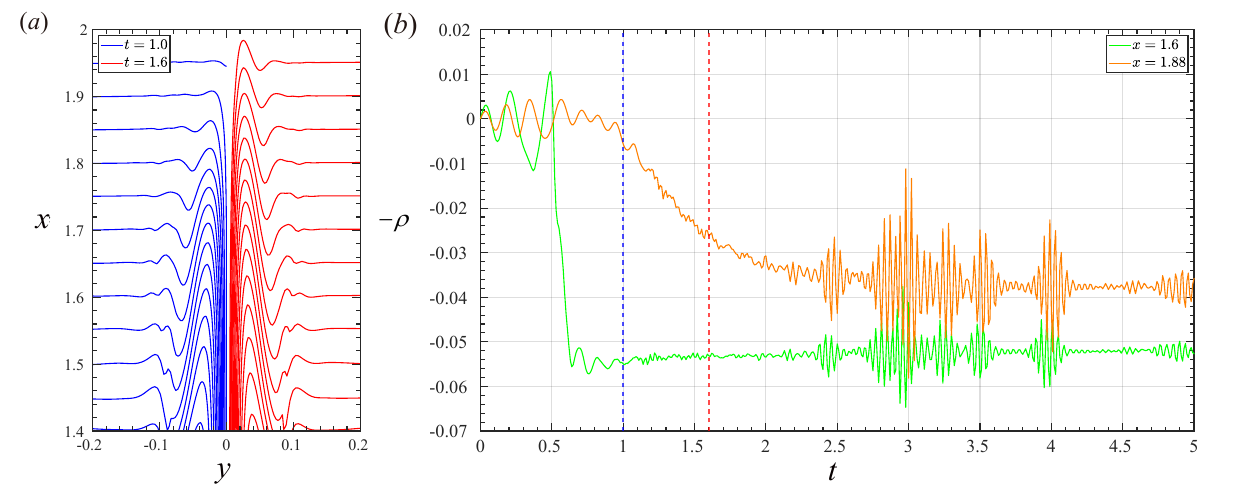}
\caption{Density field in the wake region. $(a)$: iso-pycnals lines $\rho(x,r,t)-x=const.$ at two different instants of time (identified with dashed lines in $(b)$) in the axisymmetric base flow; 
$(b)$: time history of the density disturbance $\rho(x,y,z,t)$ at two vertical locations, $x = 1.60$ and $1.88$, on the vertical line $y = 0.05, z=0$.}%
\label{fig5-6-1}
\end{figure}

In flickering flames, vortex rings result from the shear induced by the upward motion of buoyant plumes \citep{zhang2021instability, zhang2024experimental}. Here, in contrast, the mechanism that generates them stems from stratification-induced differences in the density relaxation process, as figures \ref{fig5-6} and \ref{fig5-6-1} help reveal. The sequence displayed in figure \ref{fig5-6}$(a-d)$ shows that streamlines progressively stretch and pinch off, giving birth to closed recirculating eddies. 
Panel $(e)$ shows the temporal evolution of the radial velocity along a vertical line lying some distance away from the jet axis ($y = 0.05$). At the beginning of the sequence, the radial velocity remains positive whatever the vertical position, which indicates an outward flow. Then, it becomes negative over a short range of vertical positions, $1.8 < x < 1.87$, signaling a local flow reversal and the onset of a vortex ring. Figure \ref{fig5-6-1}$(b)$, which displays the evolution of the density disturbance at two different heights along the same vertical line, helps understand the cause of this reversal. While $\rho$ quickly stabilizes at the lower position ($x=1.6$), it continues to evolve over a much longer time at the upper position ($x = 1.88$). The imbalance in the density equation \eqref{equ_2-3} responsible for this longer relaxation arises from the weaker radial diffusion at $x=1.88$ than at $x=1.6$. Indeed, as figure \ref{fig5-6-1}$(a)$ shows, the isopycnals stay further apart from each other at the upper position, resulting in weaker density gradients, hence in a longer time for the density disturbance to reach a stationary value. The delayed adjustment of the upper fluid stretches the streamlines until they separate, which yields the formation of vortex rings. This scenario is consistent with the numerical observations of \cite{hanazaki2015numerical} regarding vertical variations of the various terms of \eqref{equ_2-3} along the jet axis. 
%

The above discussion makes it clear that the vortex ring formation mechanism requires strong stratification conditions (low Froude number) and weak scalar diffusivity (high Prandtl number) for the lower fluid layer to equilibrate significantly faster than the upper layer, thereby promoting the emergence of a recirculating eddy. This is why the varicose instability only occurs in a specific parameter range, namely $Fr \lesssim 0.08,\ Re \gtrsim 50,\ Pr\gtrsim70$, as the regime map displayed in figure \ref{fig5-19} will establish.

\subsection{Stage III: Onset of sinuous instability}\label{sec4.2}

\begin{figure}
\vspace{3mm}
\centering
\includegraphics[width=0.96\textwidth]{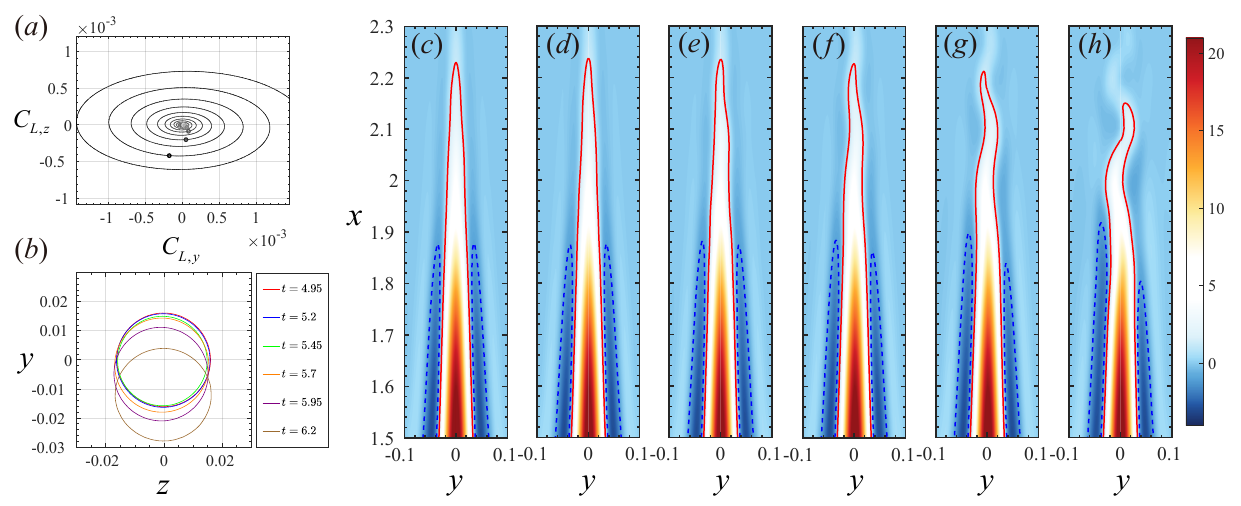}
\caption{Stage III: onset of the sinuous instability. $(a)$: trajectories of the transverse force components in the $(C_{L,y},C_{L,z})$ phase plane; $(b)$: iso-surfaces of the absolute vertical velocity $(u_x -1) = 1.2$ at $x = 2.1$; $(c-h)$: same as figure~\ref{fig5-3} $(c-h)$ in the time interval $3 < t < 6.4$; the successive snapshots are taken at times instants spotted by open circles in $(a)$.}
\label{fig5-7}
\end{figure}

As time proceeds, the varicose instability ceases and is succeeded by the sinuous instability corresponding to stage III in figure~\ref{fig5-1}$(b)$. Figure~\ref{fig5-7} summarizes this transition with the help of different indicators. The trajectories of the lift coefficients are seen to form spiral patterns in the $(C_{L,y},C_{L,z})$ phase plane (panel $(a)$), with $C_L$ growing to $\mathcal{O}(10^{-3})$, signalling the loss of axial symmetry. Similarly, panel $(b)$ shows that the centre of the iso-contours of the vertical velocity drifts away from the vertical axis $(y,z)=(0,0)$ beyond $t \approx 5$, an information confirmed by panels $(c-h)$ in which the meandering of the jet's tip is seen to gradually increase over time. The instability that sets in resembles the sinuous modes observed in diffusion flames \citep{boulanger2010laminar,zhang2021instability,xiao2023structure}. In this case, the instability arises from interactions between buoyancy and inertial forces - each dominating in different regimes - that destabilize the vortex sheet enveloping the flame, ultimately leading to the onset of the sinuous pattern.

\begin{figure}
\centering
\includegraphics[width=0.96\textwidth]{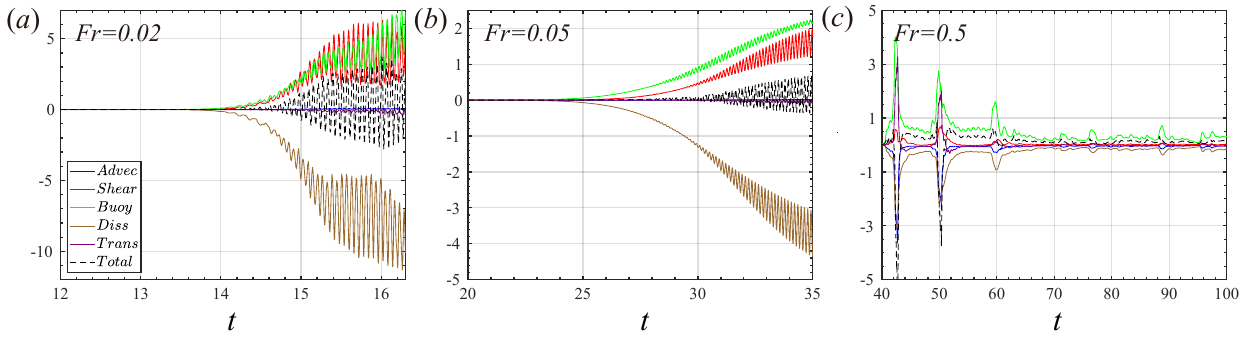}
\caption{Time evolution of the various terms in the disturbance kinetic energy budget \eqref{e5-4} at different $Fr$ for $(Re,Pr) = (100,700)$. $(a)$: $Fr = 0.02$; $(b)$: $Fr = 0.05$; $(c)$: $Fr = 0.5$.}
\label{fig5-14}
\end{figure}
To better understand the mechanisms at stake, we examine the kinetic energy budget of the non-axisymmetric disturbance, which we assume to be small in the initial stage. This budget reads \citep{lombardi2011growth,pal2017direct}
\begin{eqnarray}
\partial_t e' = && -\overline{u}_j\partial_j e' -u'_iu'_j\partial_j \overline{u}_i - Fr^{-2}\rho'u'_x -{(2Re)}^{-1}(\partial _ju'_i + \partial_i u'_j)^2 \nonumber \\
&& - \partial_i[u'_ie' + u'_ip' - 2Re^{-1} u'_j(\partial_j u'_i +\partial_i u'_j)]\,,
\label{e5-4}
\end{eqnarray}
where $e' = u_i'u'_i/2$ is the kinetic energy associated with the disturbance flow $u'_i = u_i - \overline{u}_i$ (with $\overline{u}_i$ the base flow corresponding to the axisymmetric solution), $\rho'$ and $p'$ representing the non-axisymmetric density and pressure disturbances, respectively. As mentioned in \S\,\ref{sec2}, the various terms involved in this budget are evaluated numerically by superimposing an artificial disturbance with a magnitude of $1\times10^{-4}$ onto the $x$-component of the velocity field. The first two terms in the right-hand side of (\ref{e5-4}) represent advection by the mean flow and generation by the mean shear, respectively, while the third term (sometimes referred to as the buoyancy flux) corresponds to generation by density disturbances  and the last two terms represent dissipation and transport by the disturbance, respectively. 

Figure~\ref{fig5-14} shows the time evolution of the volume-integrated terms of (\ref{e5-4}). Panels $(a-b)$ make it clear that, at low enough Froude number and sufficiently early times (up to $t\approx26$ at $Fr=0.05$), the buoyancy flux (green line) is responsible for the generation of $e'$. Generation by mean shear (red line) then starts to become significant, and both sources reach a similar magnitude after some time.  
The buoyancy flux is linked to the baroclinic vorticity generated by the tilting of velocity and density iso-surfaces near the base of the jet. The lower the diffusivity of the scalar, the most efficient this mechanism due to the strong horizontal density gradients that can then be maintained. Thus, increasing $Fr$ or decreasing $Pr$ reduces this contribution. Similarly, the shear-induced generation term decreases with increasing $Fr$ or decreasing $Pr$ since the jet then weakens and becomes thicker. These variations are confirmed in figure~\ref{fig5-14}. Indeed, according to panels $(a-b)$, both generation terms remain significant at $Fr = 0.05$, although their magnitude is smaller than at $Fr = 0.02$. Then, at $Fr = 0.5$ (panel $(c)$), they are insufficient to support the growth of the disturbance beyond some initial transient, so that the flow remains axisymmetric.
We shall examine in more detail the mechanisms responsible for the onset of the sinuous instability in \S\,\ref{sec5}.

\subsection{Stages IV and V: Transition and saturation of the sinuous instability}\label{sec4.3}

\begin{figure}
\vspace{3mm}
\centering
\includegraphics[width=0.96\textwidth]{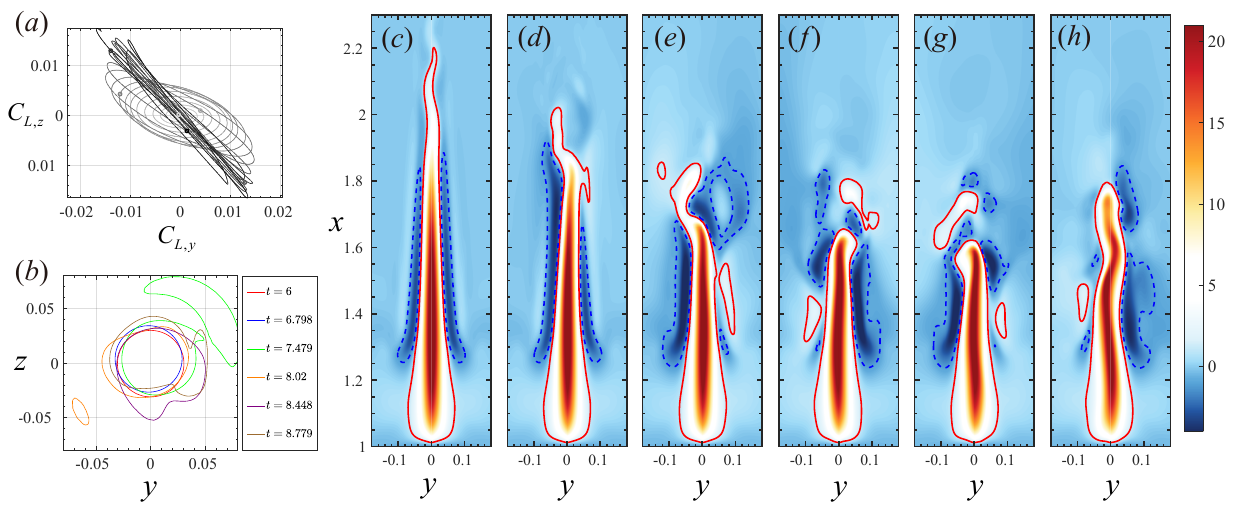}
\caption{Stage IV: gradual saturation of the sinuous instability. $(a)$: trajectories of the transverse force components in the $(C_{L,y},C_{L,z})$ phase plane; $(b)$: iso-surfaces of the vertical velocity $(u_x -1) = 1.2$ at $x = 2.0$; $(c-h)$: same as figure~\ref{fig5-3} $(c-h)$ in the time interval $6 < t < 8.8$.}
\label{fig5-15}
\end{figure}

During stage IV, \textit{i.e.}, $6 \leq t < 9$ in figure~\ref{fig5-1}$(b)$, the kinetic energy associated with the azimuthal velocity continues to increase but nonlinear effects makes its growth deviate gradually from the previous exponential trend. Simultaneously, the height of the jet drops from $H^\text{max} = 2.2$ to $\approx1.7$, as figure~\ref{fig5-15} $(c)$ reveals. Panel $(a)$ in the same figure shows that the trajectories of the transverse force components evolve from a flattened spiral to a planar zigzag as the jet shortens, indicating that the sinuous motion becomes increasingly confined to a vertical plane. 
Meanwhile, the beatings at the top of the jet reach a large amplitude, yielding strongly non-isotropic distributions of the vertical velocity in the horizontal plane (panel $(b)$). These beatings are responsible for the loss of coherence of the top part of the jet, and thus for its shortening. 

\begin{figure}
\vspace{3mm}
\centering
\includegraphics[width=0.96\textwidth]{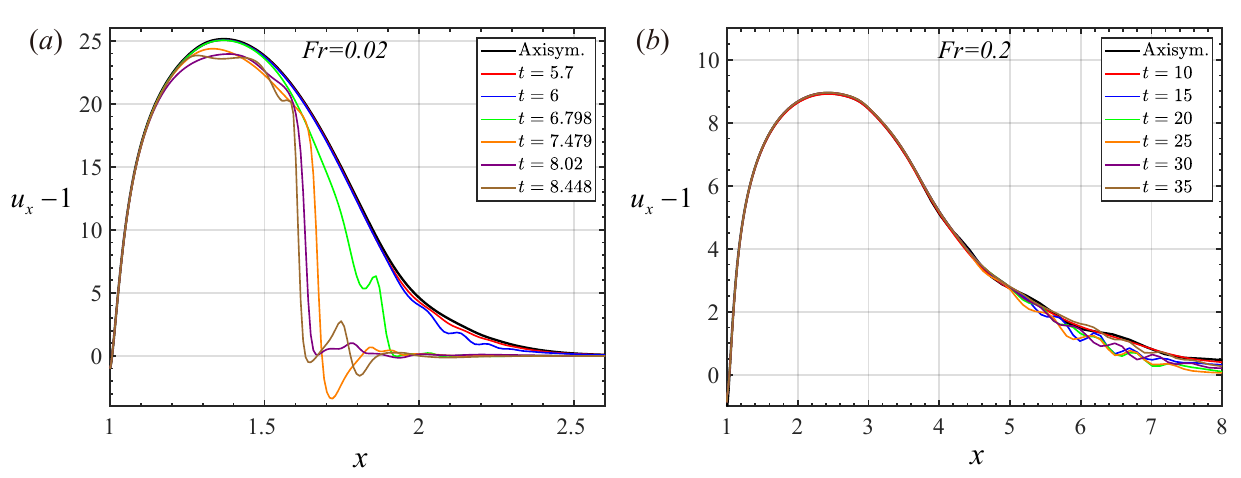}
\caption{Evolution of the absolute vertical velocity, $u_x - 1$, along the vertical axis. $(a)$: $Fr = 0.02$; $(b)$: $Fr = 0.2$.}
\label{fig5-16}
\end{figure}

The gradual saturation process at work also alters the evolution of the vertical velocity distribution along the vertical axis, as illustrated in figure~\ref{fig5-16}$(a)$. Variations in this distribution become visible at the jet's tail around $t = 5.7$ and then propagate upstream, leading to oscillations at positions $2\lesssim x \lesssim2.3$. These oscillations grow over time, leading to the breakdown of the upper portion of the jet, characterized by near-zero values of $u_x $, beyond $t = 6.8$. For comparison, figure~\ref{fig5-16}$(b)$ displays the counterpart of this evolution for a ten times larger Froude number, $Fr = 0.2$. In this case, the $u_x$-perturbations remain localized in the tail region ($x \geq 5$) over a much longer period of time ($10 \leq t \leq 35$), without significantly affecting the overall structure of the jet. 

\begin{figure}
\centering
\includegraphics[width=0.96\textwidth]{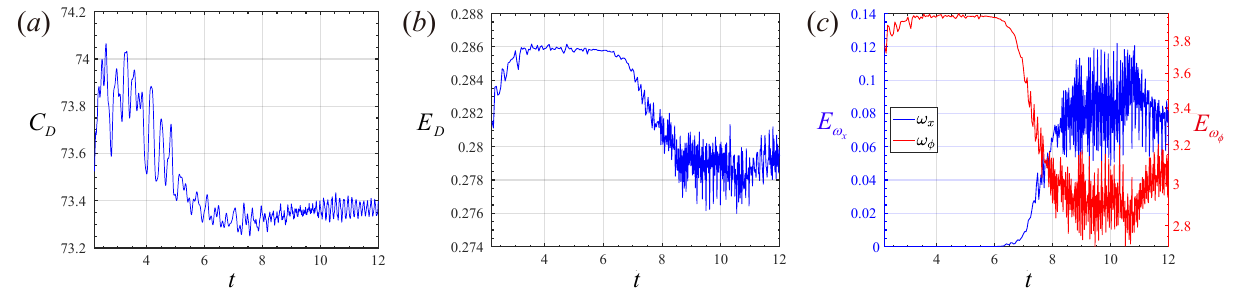}
\caption{Time history of some flow quantities during stage IV. $(a)$: drag coefficient; $(b)$: dissipation rate; $(c)$: streamwise and azimuthal enstrophy components.}
\label{fig5-17}
\end{figure}

Some consequences of the jet retraction are made apparent in figure~\ref{fig5-17}, starting with the evolution of the drag coefficient in panel $(a)$. A slight decrease of $C_D$ is observed, indicating that slightly less energy is required to sustain the sphere motion when the jet weakens. This is corroborated by the time evolution of the viscous dissipation, $E_\Phi = 2Re^{-1}\int_{\Omega} \left( \partial_j u_i  + \partial_i u_j  \right)^2 d\Omega$, shown in panel $(b)$. This quantity is seen to decrease by a few percent during this stage, owing to the reduction of the size of the high-shear region formed by the jet. 
Last, the time evolution of the enstrophy associated with the streamwise vorticity component, $E_{\omega_x} = \int_{\Omega} \omega_x^2 d\Omega$, and its counterpart for the azimuthal component, $E_{\omega_\phi} = \int_{\Omega} \omega_\phi^2 d\Omega$, are shown in panel $(c)$. Starting from near-zero values, $E_{\omega_x}$ is seen to increase sharply during stage IV, while $E_{\omega_\phi}$ simultaneously decreases, indicating a conversion of the primary azimuthal vorticity, $\omega_\phi$, into the streamwise vorticity component, $\omega_x$, through a vortex tilting mechanism (see \S\,\ref{sec5.1}).

\begin{figure}
\vspace{5mm}
\centering
\includegraphics[width=0.96\textwidth]{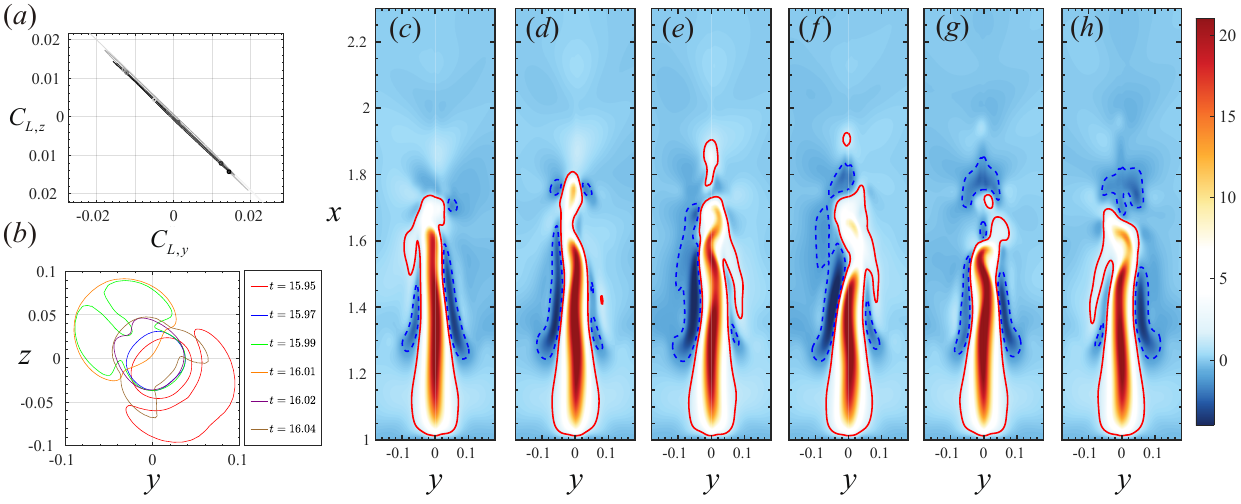}
\caption{Stage V: saturation of the sinuous instability. $(a)$: trajectories of the transverse force coefficients in the $(C_{L,y},C_{L,z})$ phase plane; $(b)$: iso-surfaces of the absolute vertical velocity $(u_x -1) = 1.2$ at $x = 1.5$; $(c-h)$: same as figure~\ref{fig5-3} $(c-h)$ in the time interval $15.95 < t < 16.04$ corresponding to one period of the jet oscillations.}
\label{fig5-18}
\end{figure}

The sinuous instability reaches a saturated state in stage V. The corresponding flow characteristics are illustrated in figure~\ref{fig5-18}. Both the evolution of the transverse force components in the $(C_{L,y},C_{L,z})$ plane (panel $(a)$) and the iso-surfaces of the vertical absolute velocity (panel $(b)$) indicate a planar zigzagging oscillation in a vertical plane close to the $y = z$ diagonal. 
Unlike in stage IV, the height of the jet remains nearly constant throughout the sequence displayed in panel $(c)$, confirming that the sinuous instability has reached saturation, leading to a statistically stationary wake structure. 

\section{Discussion}
\label{sec5}

In \S\,\ref{sec4}, we analyzed the transformation of the jet morphology through five successive stages for the parameter set $(Fr,Re, Pr = (0.02,100, 700)$. However, some of these stages do not show up during the development of the meandering jet when the control parameters are varied. For instance, when $Fr \geq 0.08$, the varicose instability does not occur, the jet transitioning directly to a non-axisymmetric oscillatory state. This observation confirms that the varicose and sinuous instabilities correspond to distinct modes in the sense of linear stability, with no direct causal relationship between them. Additionally, we observed that stage IV becomes less prominent as the Froude number increases, and virtually  disappears when $Fr \geq 0.1$. In this section, we focus on the mechanisms that trigger the sinuous instability and on the flow conditions under which this instability occurs. 
\begin{figure}
\vspace{5mm}
\centering
\includegraphics[width=1\textwidth]{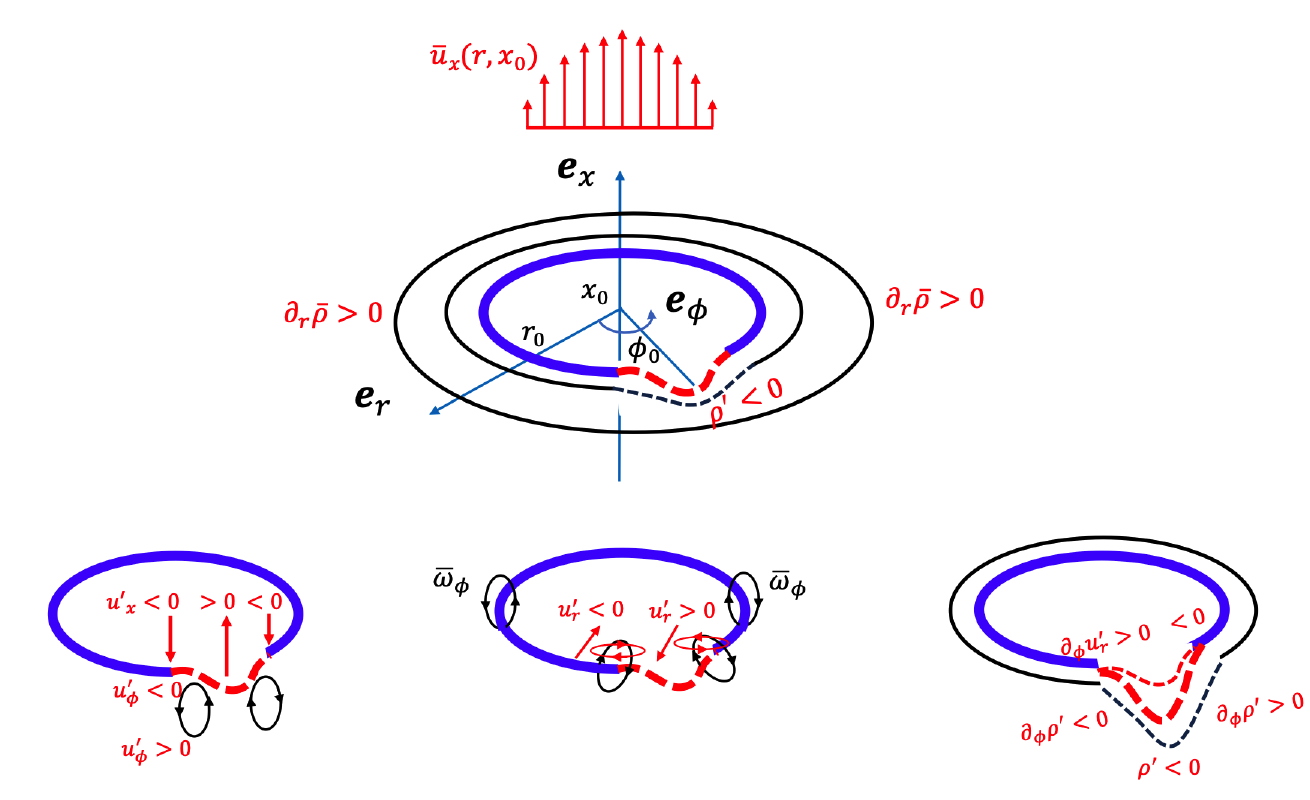}
\vspace{-65mm}\\
\hspace{-80mm}$(a)$\\
\vspace{50mm}
\hspace{-40mm}$(b)$\hspace{40mm}$(c)$\hspace{45mm}$(d)$\\
\vspace{5mm}
\caption{Sketch of the baroclinic instability responsible for the sinuous mode. $(a)$: axial velocity profile in the jet and cross section of some isopycnals in the horizontal plane $x=x_0$. A negative, $\phi$-dependent, density disturbance (red dashed tongue) induces a baroclinic torque deflecting the blue isopycnal towards a vertical position $x<x_0$. $(b)$: The baroclinic torque induces a radial vorticity disturbance, $\omega'_r$, involving a $x$-dependent azimuthal velocity disturbance, $u'_\phi$, and a $\phi$-dependent axial velocity disturbance, $u'_x$; $(c)$: the tilting of the primary vorticity, $\overline{\omega}_\phi$, by the vertical gradient of $u'_\phi$ yields an axial (vertical) vorticity disturbance, $\omega'_x$,  involving a $r$-dependent azimuthal velocity disturbance and a $\phi$-dependent radial velocity disturbance; $(d)$: reinforcement of the density disturbance $\rho'<0$ through the transport of the positive radial density gradient $\partial_r\overline{\rho}$ by the $\phi$-dependent radial velocity disturbance. }
\label{sketch}
\end{figure}
\subsection{Physical mechanism of the sinuous instability}\label{sec5.1}
\label{sec5.1}
To get some insight into the mechanism underlying the sinuous instability, we need to consider how density disturbances induce three-dimensional flow disturbances and may be reinforced by them. For this purpose, we decompose the velocity $\boldsymbol{u}$, vorticity $\boldsymbol{\omega}$ and density departure $\rho$ into their base (time-independent) component that depends only on the radial and axial coordinates, and three-dimensional time-dependent disturbances. Therefore, we write $\boldsymbol{u}(\boldsymbol{x},t)=\overline{u}_r(r,x){\boldsymbol{e}}_r+\overline{u}_x(r,x){\boldsymbol{e}}_x+\boldsymbol{u}'(\boldsymbol{x},t)$, $\boldsymbol{\omega}(\boldsymbol{x},t)=\overline{\omega}_\phi(r,x){\boldsymbol{e}}_\phi+\boldsymbol{\omega}'(\boldsymbol{x},t)$ and $\rho(\boldsymbol{x},t)=\overline{\rho}(r,x)+\rho'(\boldsymbol{x},t)$, with $\boldsymbol{u}'=(u'_r,u'_\phi,u'_x)$ and $\boldsymbol{\omega}'=(\omega'_r,\omega'_\phi,\omega'_x)$. We assume small disturbances, \textit{i.e.}, $||\boldsymbol{u}'||/(\overline{u}_r^2+\overline{u}_x^2)^{1/2}\ll1$, $||\boldsymbol{\omega}'||/|\overline{\omega}_\phi|\ll1$, $|\rho'|/|\overline{\rho}|\ll1$. Moreover, given the spatial structure of the jet and its close surroundings, we assume $|\partial_r \overline{u}_x|\gg (|\partial_x\overline{u}_x|,|\partial_r\overline{u}_r|)\gg|\partial_x\overline{u}_r|$. The first inequality is fully supported by numerical data: for instance, figures \ref{fig5-7} and \ref{fig5-16} indicate that $|\partial_r\overline{u}_x|$ is more than one order of magnitude larger than $|\partial_x\overline{u}_x|$.  Note that neglecting $|\partial_x\overline{u}_r|$ with respect to $|\partial_r\overline{u}_x|$ implies $\overline{\omega}_\phi\approx-\partial_r \overline{u}_x$. We consider the inviscid non-diffusive limit $Re\rightarrow\infty,\,Pe\rightarrow\infty$. In that limit, and with the above assumptions and approximations, the linearized equations governing the radial and axial vorticity disturbances, $\omega'_r=r^{-1}\partial_\phi u'_x-\partial_x u'_\phi$ and $\omega'_x=r^{-1}[\partial_r (ru'_\phi)-\partial_\phi u'_r]$, and the transport equation for the azimuthal derivative of the density disturbance, $\partial_\phi\rho'$, read
\indent 
\begin{eqnarray}
\label{omegar}
\overline{D}_t \omega'_r &\approx& 
r^{-1}\overline{\omega}_\phi (\partial_\phi u'_r-u'_\phi) - Fr^{-2}r^{-1}\partial_\phi\rho'\,, \\ 
\label{omegax}
\overline{D}_t\omega'_x &\approx& 
\overline{\omega}_\phi\partial_xu'_\phi\,, \\
\label{rhophi}
\overline{D}_t \partial_\phi\rho'&\approx& \partial_\phi u'_x(1-\partial_x\overline{\rho})- \partial_\phi u'_r\partial_r\overline{\rho}\,,
\label{e5-1}
\end{eqnarray}
with $\overline{D}_t=\partial_t+\overline{u}_r\partial_r+\overline{u}_x\partial_x$. Now, suppose that, at vertical and radial positions $x_0$ and $r_0$, the density slightly decreases locally around an azimuthal position $\phi_0$ (figure \ref{sketch}$(a)$). This amounts to assuming that the isopycnal crossing the horizontal plane $x=x_0$ at the radial position $r=r_0$ is locally deflected towards the sphere, since $\partial_\phi\rho'<0$ (resp. $\partial_\phi\rho'>0$) for $\phi\lesssim\phi_0$ (resp. $\phi\gtrsim\phi_0$).
Starting from an axisymmetric flow with $\omega'_r=\omega'_x=0$, this disturbance creates a positive $\omega'_r$ at angular positions $\phi\lesssim\phi_0$ according to \eqref{omegar}, hence a positive $\partial_\phi u'_x$ and a negative $\partial_x u'_\phi$ according to the definition of $\omega'_r$ (figure \ref{sketch}$(b)$). Since the axial velocity $\overline{u}_x$ decreases with increasing $r$ in the jet, $\overline{\omega}_\phi$ is positive, so that the right-hand side of \eqref{omegax} becomes locally negative for $\phi\lesssim\phi_0$, yielding a negative axial vorticity disturbance $\omega'_x$ (figure \ref{sketch}$(c)$). This in turn implies $\partial_r(ru'_\phi)<0$ and $\partial_\phi u'_r>0$, which provides a second positive contribution to the right-hand side of  \eqref{omegar} through the vortex tilting term $r^{-1}\overline{\omega}_\phi \partial_\phi u'_r$.\\
\indent To see how the local velocity gradients $\partial_\phi u'_x$ and $\partial_\phi u'_r$ resulting from $\omega'_r$ and $\omega'_x$ affect the density disturbance, it is first worth noting that $1-\partial_x\overline{\rho}=\partial_x(x-\overline{\rho})$ and $-\partial_r\overline{\rho}=\partial_r(x-\overline{\rho})$. Since any isopycnal in the base flow obeys the equation $x-\overline{\rho}(r,x)=x_\infty$, with $x_\infty$ denoting the vertical position of this isopycnal far from the sphere, \textit{i.e.}, for $r\rightarrow\infty$, $\partial_x(x-\overline{\rho})$ and $\partial_r(x-\overline{\rho})$ are nothing but the axial and radial derivatives of $x_\infty$.
Therefore, \eqref{rhophi} may be recast in the form 
\begin{equation}
\label{recast}
\overline{D}_t \partial_\phi\rho'\approx \partial_\phi u'_x\partial_xx_\infty+ \partial_\phi u'_r\partial_rx_\infty\,.
\end{equation}
Let us first assume that the $\phi$-dependent density disturbance occurs in the core of the jet, \textit{i.e.}, $r_0\ll1$. In this thin region, isopycnals are almost vertical when $Fr$ is small, and density increases radially outwards, as figure \ref{fig5-6-1}$(a)$ depicts. Therefore, $\partial_xx_\infty\approx0$ and $\partial_rx_\infty<0$, which makes the right-hand side of \eqref{recast} negative at positions $\phi\lesssim\phi_0$, yielding $\overline{D}_t \partial_\phi\rho'<0$. Hence, owing to the shape of the isopycnals within the jet, the density balance reinforces the negative azimuthal gradient of the initial density disturbance (figure \ref{sketch}$(d)$). The reasoning remains unchanged at positions $\phi\gtrsim\phi_0$ where $\partial_\phi\rho'>0$. Thus, the axisymmetric jet undergoes a baroclinic instability that tends to make it fully three-dimensional. Obviously, the smaller $Fr$ the larger the source term in \eqref{omegar}, hence the stronger the efficiency of the instability mechanism. Let us now consider slightly larger $r_0$ corresponding to radial positions lying in the close surroundings of the jet. As $r$ increases, isopycnals first reach a maximum altitude, before coming back down and eventually returning to their equilibrium position after some oscillations (see figure \ref{fig5-6-1}$(a)$). In this outer region, say for $r>0.03$ in figure \ref{fig5-6-1}$(a)$, 
$\partial_rx_\infty$ is either close to zero or positive, whereas $\partial_xx_\infty$ is positive everywhere. Therefore, the right-hand side of \eqref{recast} is now positive at angular positions $\phi\lesssim\phi_0$, and so is $\overline{D}_t \partial_\phi\rho'$. This reduces the magnitude of the initial disturbance $\partial_\phi\rho'<0$, contributing to make the density field surrounding the jet return to its axisymmetric base state.  \\
\indent Obviously, viscous and diffusive effects neglected in this qualitative analysis tend to damp $\omega'_r$, $\omega'_x$ and $\partial_\phi\rho'$ and are presumably able to restore the stability of the jet below some critical, $Fr$-dependent, value of the Reynolds and P\'eclet numbers. Only a rigorous stability analysis solving the generalized eigenvalue problem associated with three-dimensional disturbances superimposed on the axisymmetric $(Fr,Re,Pe)$-dependent base state can provide the actual bounds of the unstable domain in the parameter space, as well as the frequency and spatial structure of the most unstable (or least stable) eigenmode.\\
\begin{figure}
\vspace{4mm}
\centering
\includegraphics[width=0.96\textwidth]{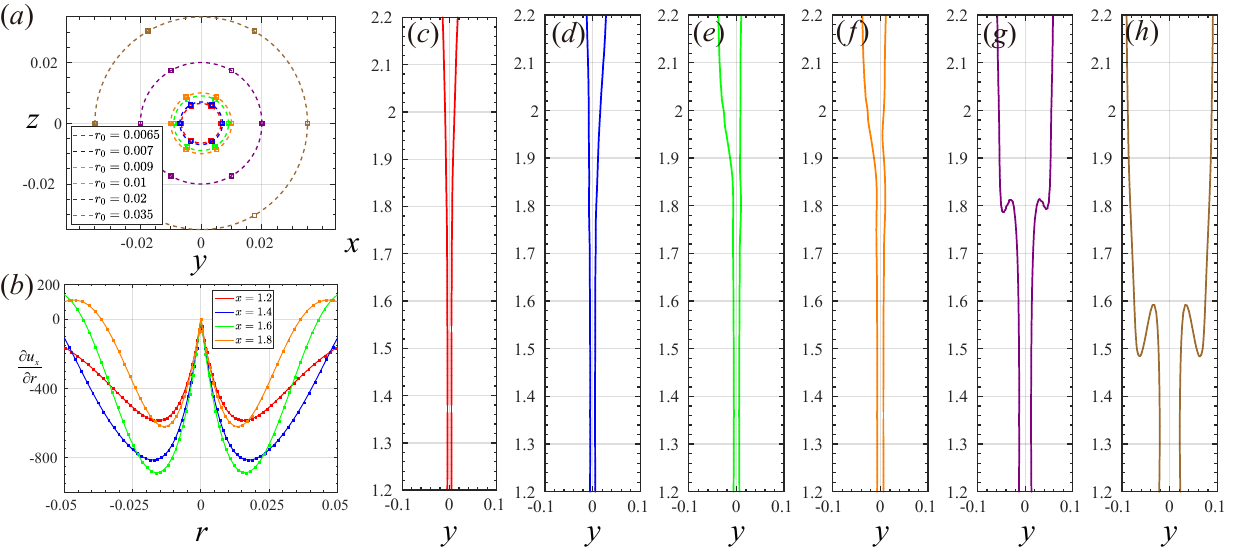}
\caption{Trajectories of Lagrangian particles released at the basis of the jet at $t=4$. 
$(a)$: schematic of the initial particle positions, all released at the axial location $x=1.0$ but with varying radial positions $0.0065 \le r_0 \le 0.035$ and equidistant azimuthal positions differing by an angle $\Delta\phi = \pi/3$; $(b)$: radial profiles of the main velocity gradient $\partial_r \overline{u}_x$ at different axial positions; $(c-h)$: projections in the vertical $(x,y)$ plane of trajectories of two particles initially placed at the same $r_0$ and respective azimuthal positions $\phi = 0$ and $\pi$ ($r_0$ is specified in $(a)$ and increases from left to right in $(c-h)$). 
}
\label{fig5-10}
\end{figure}
\begin{figure}
\vspace{4mm}
\centering
\includegraphics[width=0.96\textwidth]{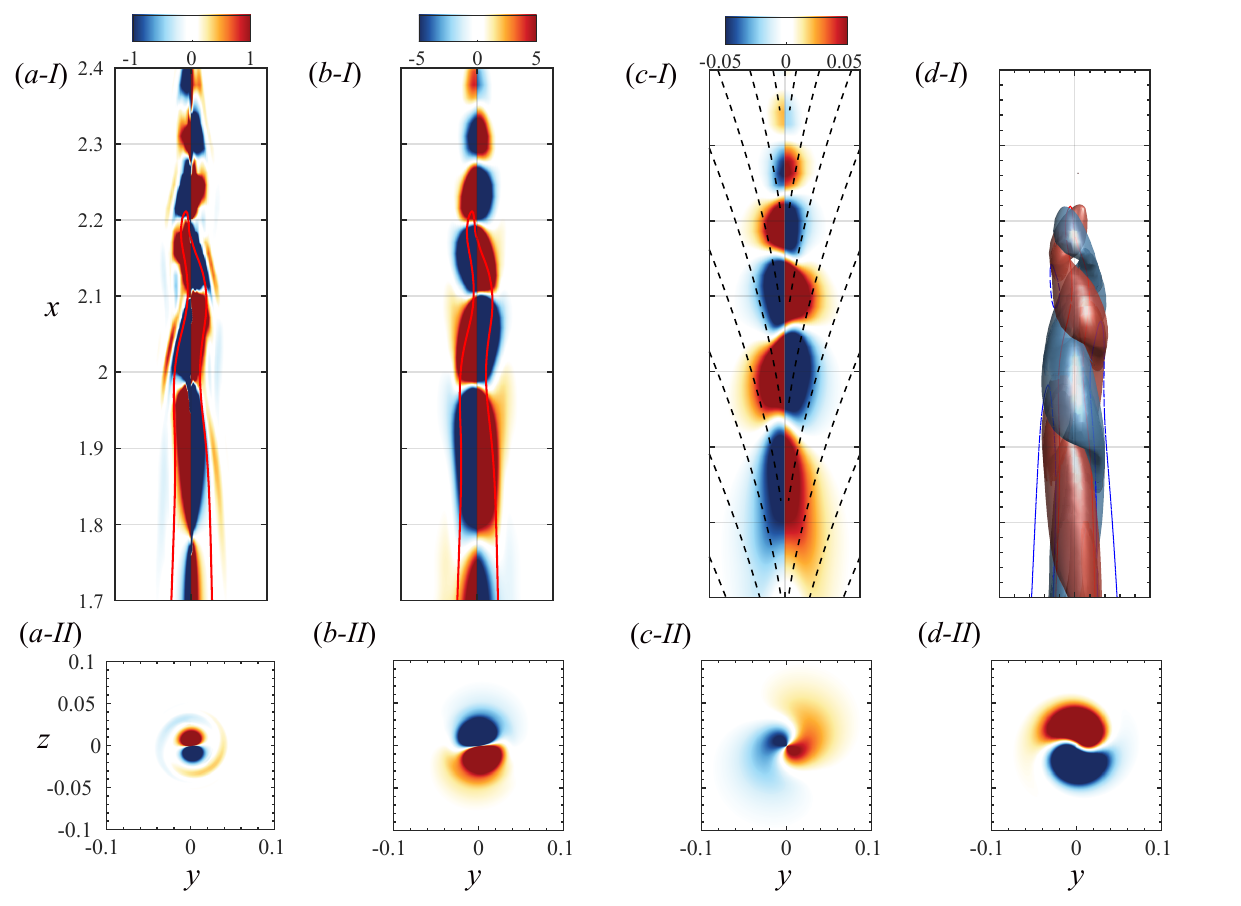}
\caption{Some features of the wake structure in the presence of the sinuous instability at the end of stage III ($t=5.95$).$(a-I)$: iso-values of the radial component of the baroclinic torque, $r^{-1}Fr^{-2} \partial_\phi\rho$, in the vertical diametrical plane $z=0$ (the iso-contour $u_x-1=1.2$ is shown with a thin red line); $(b-I)$: same for  the radial vorticity component, $\omega_r$; $(c-I)$: same for the azimuthal velocity, $u_\phi$, with dashed lines representing the lee wave pattern defined by $u_x - 1 = 0$; $(d-I)$: 3D iso-surfaces $\omega_x=\pm1$ of the axial (streamwise) vorticity component. Row $II$: same quantity as in the corresponding panel in row $I$ at the vertical position $x=1.8$.}
\label{fig5-9_new}
\end{figure}
\indent To better identify the flow region responsible for the growth of the sinuous instability, we released Lagrangian particles at various azimuthal positions from the rear of the sphere, very close to the vertical axis, at $t=4$, \textit{i.e.}, just at the beginning of stage III (figure~\ref{fig5-10}$(a)$). 
At this time, the main shear $\partial_r \overline{u}_x$ peaks at $r\approx 1.5\times10^{-2}$ (figure~\ref{fig5-10}$(b)$). Panels $(c-h)$ show how two particles released symmetrically at angular positions $\phi = 0$ and $\phi=\pi$ at the same radial position $r_0$ from the jet axis evolve. The pair released closest to the axis (panel ($c$)) rises outwards almost symmetrically, although a slight asymmetry may be discerned in the top region ($x\gtrsim2.0$). The three pairs in panels $(d-f)$, released at radial positions $0.007 \le r_0 \le 0.01$, exhibit markedly asymmetric paths beyond $x \approx 1.8$. At this vertical position, their radial position is close to $r=0.015$,  which corresponds to the maximum of the shear rate according to panel $(b)$. In contrast, in panels $(g-h)$, corresponding to  $r_0 \ge 0.02$, trajectories are seen to preserve a mirror symmetry throughout their course. Now they reach the position $x=1.8$ (panel $(g)$) or $x=1.6$ (panel $(h)$) at $r\gtrsim0.03$, where the shear has been reduced to about one-third of its maximum value. Beyond this point, they exhibit a back-and-forth oscillation caused by the bell-shaped flow structure around the jet \citep{hanazaki2009jets, hanazaki2015numerical}. 
 Thanks to this ``capture-release" phenomenon, particles escape from the jet and reach a weakly sheared region. The strikingly different fate of  particle trajectories in panels $(d-f)$ as compared to panels $(g-h)$ confirms the crucial role of the main shear $\partial_r\overline{u}_x$ (hence, of $\overline{\omega}_\phi$) in the instability mechanism. If this shear only reaches moderate levels (as is the case in the jet's immediate surroundings), the tilting mechanism in \eqref{rhophi} only produces a modest azimuthal gradient of $u'_r$. Moreover, the density gradient  $\partial_r\overline{\rho}$ is much weaker there than in the core of the jet. Therefore, the right-hand side of \eqref{recast} is dominated by the stabilizing contribution $\partial_\phi u'_x\partial_xx_\infty$, keeping the flow locally axisymmetric. Conversely, at radial positions where $|\omega_\phi|$ is close to its maximum, the tilting mechanism produces much larger $\partial_\phi u'_r$ which, combined with the large positive radial density gradient, makes the source term $\partial_\phi u'_r\partial_rx_\infty$ dominant in the right-hand side of \eqref{recast}, leading to the growth of the sinuous instability.  
\\
\indent Some aspects of the flow structure resulting from the baroclinic instability described above may be appreciated at a slightly later stage in figure \ref{fig5-9_new}. As the bottom row of the figure makes clear, the dominant emerging three-dimensional mode is associated with an azimuthal wavenumber $m=\pm1$ (one period corresponds to a $2\pi$-variation of $\phi$). In the vertical direction, this mode exhibits a specific distribution. Namely, at a given time, the azimuthal density gradient (panel $(a-I)$), radial vorticity (panel $(b-I)$), and azimuthal velocity (panel $(c-I)$) exhibit an alternation of positive and negative values along the jet axis, the height of each ``spot" being a decreasing function of the downstream position with respect to the sphere. The axial vorticity pattern in panel $(d-I)$ consists of a pair of twisted threads. Such helical structures are consistent with the spiralling pattern of the $(C_{L, y},C_{L, z})$ trajectories observed in figure~\ref{fig5-7}$(a)$. The dashed lines in panel $(c)$, which correspond to iso-lines $u_x=1$, suggests that the internal waves radiated by the sphere have a direct influence on the axial modulation and radial extension of the non-axisymmetric flow components. Note that the wake structure in figure \ref{fig5-9_new} is transitional. Later, it gradually switches to the standing wave mode structure displayed in figure \ref{fig4-3}$(c-d)$.   

\begin{figure}
\centering
\includegraphics[width=0.96\textwidth]{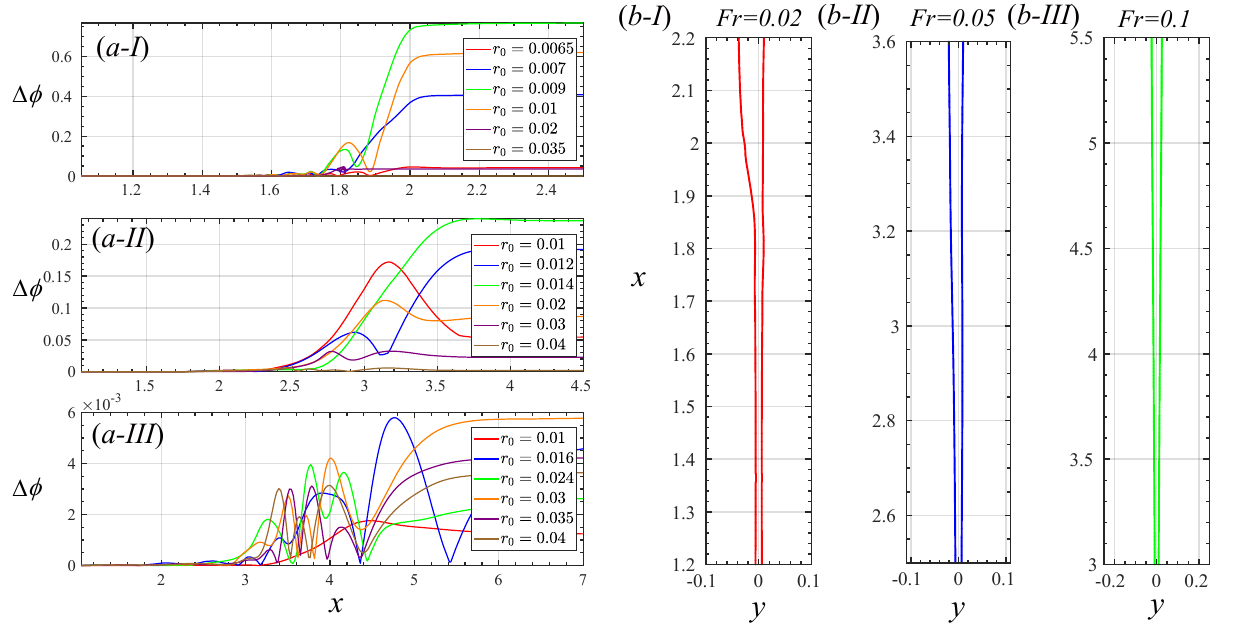}
\caption{Influence of the Froude number on the evolution of fluid particle trajectories. $(a)$: evolution of the azimuthal deviation, $\Delta\phi$, for particles released at various radial positions $r_0$ and at the angular position $\phi_0=0$. Panels (I-III) correspond to $Fr=0.02$, $0.05$, and $0.1$, respectively; $(b)$: trajectories of particles with the largest $\Delta\phi$ identified at each Froude number, namely $r_0=0.009$, $0.014$ and $0.03$ for increasing $Fr$, as highlighted in $(a)$.}
\label{fig5-12}
\end{figure}


\subsection{Influence of control parameters on the sinuous instability}\label{sec5.2}
Previous investigations established that the three control parameters, $Fr,\,Re,\,Pr$ have a crucial influence on the flow structure and stability. To better quantify the impact of the density stratification at low Froude number, we use again the Lagrangian tracking technique to explore $Fr$-induced changes on the motion of particles released at the top of the sphere, keeping the Reynolds and Prandtl numbers unchanged ($Re = 100,\,Pr = 700$). 
Starting from the same arrangement as in figure \ref{fig5-10}$(a)$, figure~\ref{fig5-12} summarizes the results obtained for $Fr = 0.02$, $0.05$, and $0.1$. Panels $(a)$-I to $(a)$-III display the evolution of the azimuthal deviation, $\Delta\phi$, of particles released at various $r_0$ but at the same angular position $\phi_0$ for the three stratification levels. As $Fr$ increases, the magnitude of $\Delta \phi$ decreases dramatically, with the maximum deviation in each series reducing from $\Delta \phi_{\max} = 0.76$ at $Fr = 0.02$ to $\Delta \phi_{\max} = 5.6 \times 10^{-3}$ at $Fr = 0.1$. Panels $(b)$-I, to $(b)$-III show the path of particles with the largest $\Delta \phi$ at each $Fr$. With no surprise, the larger $Fr$ the more symmetric the path, consistent with a decrease in the amplitude of the unstable sinuous mode. Interestingly, the initial radial position leading to the largest $\Delta\phi$ shifts outwards with increasing $Fr$, from $r_0 = 0.009$ at $Fr = 0.02$ to $r_0 = 0.03$ at $Fr = 0.1$. This variation aligns with the expected scaling of the jet radius, which grows as $Fr^{1/2}$ \citep{hanazaki2015numerical, okino2023schmidt}. 

\begin{figure}
\centering
\includegraphics[width=0.96\textwidth]{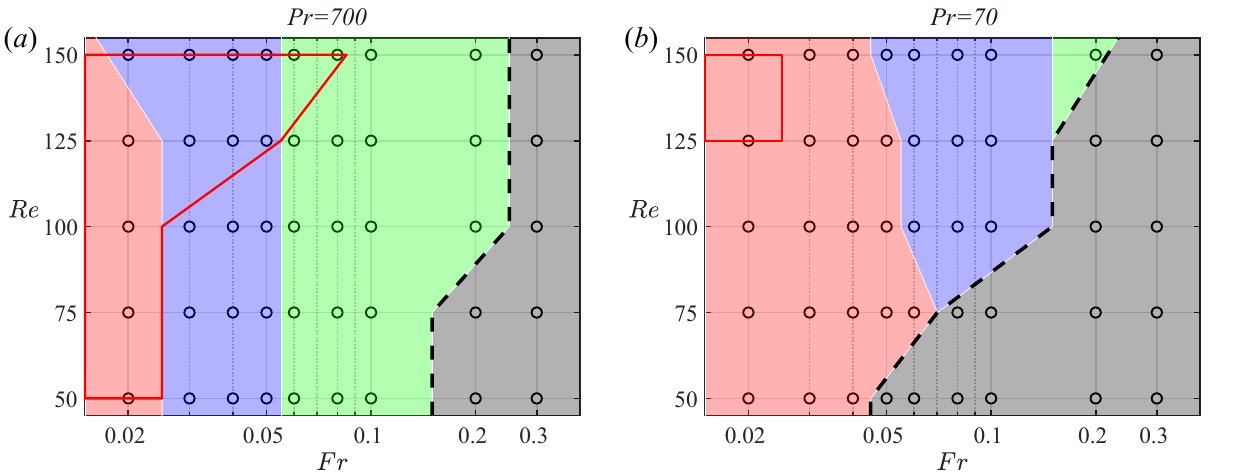}
\caption{State diagram summarizing the flow regimes encountered in the late stage of the simulations. $(a)$: $Pr=700$; $(b)$: $Pr=70$. Grey regions correspond to a stable axisymmetric jet, whereas each colored region corresponds to a specific unstable configuration identified through the $(C_{L,y},C_{L,z})$ diagram, with green, purple and pink referring to chaotic, spiral and standing wave modes, respectively. The red contour delineates the sub-region exhibiting a transient varicose instability prior to the onset of the sinuous instability.}
\label{fig5-19}
\end{figure}


To further explore the influence of the control parameters, we ran a large number of simulations based on different combinations of $(Fr,\,Re,\,Pr)$. Figure~\ref{fig5-19} synthesizes the corresponding findings obtained in the fully-developed stage in the form of a regime map covering the parameter range $0.02 \leq Fr \leq 0.3$, $50 \leq Re \leq 150$, for both $Pr = 70$ and $700$. 
We also considered the lower value $Pr = 7$ typical of heat diffusion in water. However, we could only identify a few $(Fr,\,Re)$ combinations exhibiting a sinuous instability, all for $Fr =0.02$ and $Re \geq 100$.

In figure~\ref{fig5-19}$(a)$ ($Pr = 700$), the transition separating stable and unstable configurations (black dashed line) is observed to lie in the range $0.1 < Fr < 0.2$ for $50 \leq Re \leq 75$, and shifts upwards to $0.2 \lesssim Fr \lesssim 0.3$ for $Re \geq 100$. This is consistent with the intuitive idea that the weaker the viscous diffusion, the more prone the jet is to the sinuous instability. As $Pr$ is reduced by a factor of ten (panel $(b)$), the critical $Fr$ beyond which the jet remains stable becomes more sensitive to viscous effects, reducing from $0.2 \lesssim Fr \lesssim 0.3$ at $Re=150$ to $0.04 \lesssim Fr \lesssim 0.05$ at $Re = 50$. Moreover, the critical $Fr$ at a given $Re$ is seen to shift towards lower values, especially when $Re<100$. This trend arises because diffusive and viscous effects cooperate to reduce both the main shear $\partial_r\overline{u}_x$ and the radial density gradient $\partial_r\overline\rho$, both of which play a key role in the instability mechanism detailed in \S\,\ref{sec5.1}. Similar behaviour has been reported in the linear stability analysis of planar thermal plumes, where sinuous unstable modes emerge at high $Pr$ ($Pr > 100$) and sufficient Grashof numbers but are absent otherwise \citep{lakkaraju2007effects}.

Additionally, figure~\ref{fig5-19} highlights the existence of three distinct unstable sinuous modes, identified thanks to the behaviour of the transverse force components in the $(C_{L,y},C_{L,z})$ plane. It is observed that, for high enough $Re$, increasing $Fr$ promotes a transition from a planar standing wave mode to a spiral mode, and ultimately to a chaotic regime. Nevertheless, for $Pr=70$, no chaotic regime is observed below $Re\approx125$, and the jet even switches directly from an unstable non-axisymmetric configuration dominated by a standing wave mode to a stable axisymmetric configuration for $Re\lesssim75$. The standing wave/spiral/chaotic sequence  is consistent with observations in the wake of heated spheres \citep{kotouvc2009transition}. {Last, comparing panels $(a)$ and $(b)$ indicates that the critical $Fr$ marking the standing wave/spiral and spiral/chaotic transitions shifts towards smaller values as $Pr$ increases. In particular, the critical $Fr$ corresponding to the standing wave/spiral transition decreases from $\approx0.055$ at $Pr=70$ to $\approx 0.025$ at $Pr=700$ over most of the $Re$-range displayed in the figure. The reason for this decrease stems for the variation of the jet's length with $Pr$. Diffusive effects smoothing out density gradients, increasing $Pr$ for a given pair $(Fr,Re)$ makes the jet longer, which in turn facilitates the emergence of a fully-three dimensional wake pattern, thus favouring the spiral mode at the expense of the planar standing wave mode. 

\section{Summary and concluding remarks}
\label{sec6}

In this study, three-dimensional simulations of the flow past a rigid sphere settling through a fluid with a strong linear density stratification were carried out over a broad range of Froude, Reynolds, and Prandtl numbers. The results reveal a rich sequence of wake behaviours as buoyancy effects become increasingly pronounced. For weak-to-moderate stratifications, the wake remains essentially axisymmetric, characterized by a narrow, high-speed upward jet forming at the rear of the sphere. However, under sufficiently strong stratifications, this jet becomes unstable and eventually exhibits a sinuous, meandering motion, already revealed by experimental observations. 
The main outcome of this work is two-fold: first, the simulations capture in detail the successive stages of the growth of the jet instability; second, they validate the key role of the central ingredients of the  instability mechanism that has been proposed, namely the combined presence of a strong negative radial shear and a large positive radial density gradient within the jet. 

Five distinct stages in the evolution of the jet instability have been identified. Initially, following the formation of a vertically aligned jet, an axisymmetric ``varicose" instability emerges in highly stratified configurations ($Fr \leq 0.08$), where the jet periodically bulges and necks-in along its axis. This varicose mode arises from the interaction between the initial buoyant vortex ring at the back of the sphere and the developing jet, analogous to the flickering observed in other buoyant jets and plumes and diffusion flames. However, this mode only occurs for sufficiently large $Pr$ and small $Fr$, since it relies on conditions where the fluid layer closest to the sphere equilibrates more rapidly than the layer located further downstream in the wake, leading to streamlines separation. Subsequently (or independently), the ``sinuous" instability sets in, in which the jet loses its axial symmetry and begins to meander. During the early growth of the sinuous mode, the transverse force acting on the sphere increases to finite values, and axial (streamwise) vorticity is generated periodically along the edges of the jet. An energy budget analysis reveals that production by the mean shear and by buoyancy contribute nearly equally to the growth of the non-axisymmetric perturbation. 
At later times, a ``transitional sinuous" stage may develop under strong enough stratifications ($Fr \leq 0.1$), during which the jet oscillations become approximately confined to a two-dimensional vertical plane and propagate upstream, progressively eroding the jet's upper part until the flow enters a ``saturated sinuous" stage. In this final stage, the jet reaches a quasi-steady oscillatory state in which the fluctuations of the transverse force remain periodic with nearly constant amplitude. Analysis of the evolution of this force throughout the successive stages reveals that its trajectories in the $(C_{L,y},C_{L,z})$ plane are a good metric to qualify the wake dynamics. When the jet remains long and the instability develops far downstream from the sphere (which happens for $0.025\lesssim Fr\lesssim0.2$ at $Pr=700$), this trajectory exhibits chaotic or spiral-like transitions. In contrast, in cases where the jet becomes short and thin ($Fr<0.025$ at $Pr=700$), the trajectory exhibits a planar zig-zagging pattern.

We also investigated the mechanisms underlying the onset of the sinuous instability. 
For this purpose, we considered linearized vorticity balances in the radial and azimuthal directions together with the transport equation for the azimuthal derivative of the density disturbance, neglecting viscous and diffusive effects. We showed that, due to subtle couplings and equilibria between the relevant components of the velocity-gradient tensor, the combined effect of a $\phi$-dependent density disturbance and of the strong negative radial mean shear within the jet results in an azimuthal gradient of the radial velocity disturbance through a vortex tilting process. It turns out that, combined with the large positive radial density gradient in the jet's core, this $\phi$-dependent velocity disturbance enhances the initial density disturbance, leading to a non-axisymmetric flow structure. Conversely, beyond the jet's edge, the largest density gradient in the base flow is along the vertical direction and the same mechanism leads to a damping of the density disturbance. Therefore, the sinuous instability appears to require the concomitance of a large shear rate (hence, a sufficient Reynolds number) and a large radial density gradient (hence, a sufficient P\'eclet number), with strong enough stratification effects (hence, a low enough Froude number). Numerical results make it clear that the internal waves radiated by the sphere participate in shaping the spatial structure of the sinuous mode. Nevertheless, quantifying this aspect as well as other possible roles of internal waves deserves further investigation. 

This study provides a better understanding of how stratification fundamentally alters the wake dynamics of an axisymmetric body settling along its axis. Nevertheless, as in any DNS approach, it leaves important aspects of the early stages of the jet instability unsolved. Determining the first steps of the bifurcation sequence and how the characteristics of the base flow influence both the threshold and the frequency of the first unstable (or least stable) modes is still missing but may be explored using modern tools allowing the computation of unstable global modes in complex flows \citep{Fabre2018}. In a second step, weakly nonlinear approaches may be implemented to understand how nonlinearities induced by the first unstable modes modify the base flow and how mode coupling selects the ``style" of wake oscillations \citep{fabre2008bifurcations,auguste2009bifurcations,Tchoufag2015}.
Besides these fundamental aspects, the present investigation leaves out many crucial aspects of the problem relevant with respect to applications. In particular, in most geophysical or engineering systems, particles are not spherical, nor even axisymmetric, and their geometric anisotropy introduces additional complexities, the first of which being that their path is generally non-vertical even in regimes where the wake is still stable; see, \textit{e.g.}, \cite{mrokowska2018stratification,mercier2020settling,more2021orientation} for disks and spheroids. Extending the present work to such anisotropic bodies, as well as to deformable bodies, is an exciting objective for future studies.\\


\noindent{\bf Funding} {C.-F. M. and J. Z. gratefully acknowledge the support of the National Key R$\&$D Program of China under grant 2023YFA1011000 and of NSFC under grant W2511004. All authors acknowledge the support of Toulouse INP through the ETI Programme 2025.} \\
\vspace{1mm}\\
\noindent{\bf Declaration of Interests} {The authors report no conflict of interest.}\\
\vspace{1mm}\\
\noindent{\bf Author ORCIDs}\\
{
Chang-Fan Mo,  https://orcid.org/0009-0007-5442-9434;\\
Matthieu J. Mercier, https://orcid.org/0000-0001-9965-3316;\\
Jacques Magnaudet, https://orcid.org/0000-0002-6166-4877;\\
Jie Zhang, https://orcid.org/0000-0002-2412-3617.
}

\appendix
\section{Grid design and grid-independence tests}
\label{appA}

\begin{figure}
\centering
\includegraphics[width=0.96\textwidth]{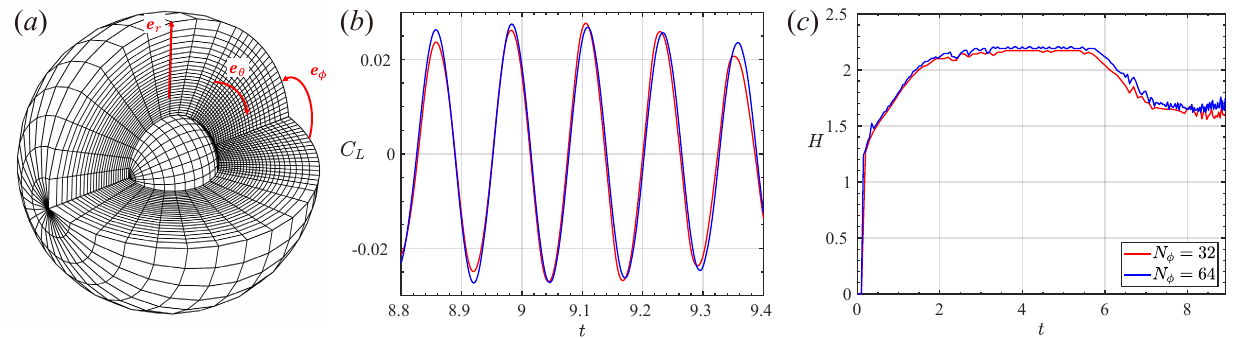}
\caption{Computational domain and azimuthal resolution effects in the case $(Fr, Re, Pr) = (0.02,100, 700)$, using a base grid with $N_r \times N_\theta = 200 \times 420$, with $\Delta_r = \Delta_\theta = 5.0 \times 10^{-4}$. $(a)$: global view of the grid for $N_\phi = 64$ with, for clarity, only one out of every ten cells is shown in the $r$- and $\theta$- directions, and one out of every two cells is shown in the $\phi$ direction; $(b)$: time evolution of the lift coefficient for $N_\phi = 32$ and $64$; $(c)$: evolution of the vertical position of the maximum jet asymmetry (as defined in figure~\ref{fig5-1}$(b)$).}
\label{fig6-2}
\end{figure}

The accuracy and efficiency of the JADIM code for simulating the stratified jet behind a settling sphere in a 2D axisymmetric framework was demonstrated by \cite{zhang2019core}. Results obtained in this earlier publication fully reproduce existing numerical data across a wide range of parameters \citep{yick2009enhanced,hanazaki2015numerical}.

In the present study, computations are still performed on a spherical grid with a non-uniform distribution of cells in both the radial ($r$) and polar ($\theta$) directions, as illustrated in figure~\ref{fig6-2}$(a)$. The setting is similar to that used in the axisymmetric configuration. More precisely, to ensure accurate resolution of the wake, the grid is divided into two regions along the $\theta$-direction. The first of them, within which the cell distribution in $\theta$ is non-uniform, corresponds to a conical subdomain centered on the upper half of the flow axis with a half-angle of $60^\circ$. The second region covers the remainder of the domain with a uniform distribution. In the first region, the polar grid spacing is progressively refined towards the upper pole using a geometric progression with a common ratio of $1.02$. In the $r$-direction, starting from the sphere surface ($r = 1$), the spacing increases outwards with a geometric ratio of $1.08$. The outer boundary is placed at $r_{\text{max}} = 40$, based on convergence tests that showed negligible sensitivity of the results when this boundary was placed at least twice as far from the sphere.

The most demanding case in this study, characterized by the weakest diffusive effects and strongest stratification ($Fr = 0.02, Re = 100, Pr = 700$), employs a grid with $200 \times 420$ cells in the $(r, \theta)$ plane. At the sphere surface, the minimum radial spacing, $\Delta_r$, and the polar spacing near the upper pole, $\Delta_\theta$, are identical, both equal to $5.0 \times 10^{-4}$. Extending this discretization to three dimensions implies rotating the $(r, \theta)$ discretization by a $2\pi$-angle in the azimuthal direction, which is discretized with 64 cells, yielding an azimuthal resolution $\Delta\phi = \pi/32$. The complete grid, sketched in figure~\ref{fig6-2}$(a)$, thus comprises $N_r \times N_\theta \times N_\phi = 200 \times 420 \times 64$ cells. 

To assess the grid influence, we employed the above case as a validation test. We first examined the influence of the azimuthal resolution $N_\phi$, considering $N_\phi = 32$ and $64$. Figure~\ref{fig6-2}$(b)$ indicates that the difference observed over time on the crest-to-crest amplitude of the lift coefficient is less than $5\%$, while there is negligible difference on the frequency. Then, recording the time history of $h^\text{max}$ (figure~\ref{fig6-2}$(c)$) shows that the predictions obtained on the two grids never differ by more than $3.5\%$, and the five distinct flow regimes are clearly identified in both cases. We did not test the finer resolution $N_\phi = 128$, since preliminary tests suggested that the computational time would then exceed one year on two AMD EPYC 7742 CPUs (each with 64 cores at 2.20 GHz and 256 GB of RAM). Given that the results are already in close agreement for $N_\phi = 32$ and $64$, we adopted $N_\phi = 64$ in all subsequent simulations.

\begin{figure}
\vspace{3mm}
\centering
\includegraphics[width=0.96\textwidth]{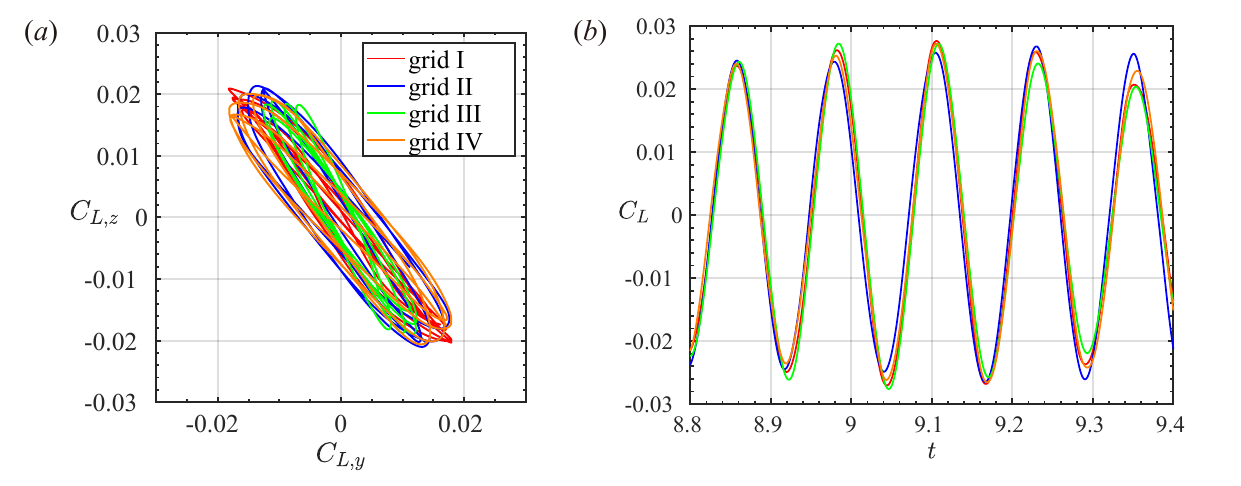}
\caption{Influence of the radial and polar resolutions on the lift coefficient in the case $(Fr, Re, Pr, Fr) = (0.02,100, 700)$. $(a)$: trajectories of the lift coefficient components in the $(C_{L,y},C_{L,z})$ phase plane; $(b)$: evolution of the total lift coefficient on four different grids: grid I ($\Delta_r = \Delta_\theta = 5 \times 10^{-4}$), grid II ($\Delta_r = 1\times10^{-3}, \Delta_\theta = 5 \times 10^{-4}$), grid III ($\Delta_r = 5 \times 10^{-4}, \Delta_\theta = 1\times10^{-3}$), and grid IV ($\Delta_r = 2.5 \times 10^{-4}, \Delta_\theta = 5 \times 10^{-4}$). 
}
\label{fig6-3}
\end{figure}

Next, we assessed the grid sensitivity in the radial and polar directions. Starting from the reference grid (hereinafter referred to as grid I) with $N_r \times N_\theta \times N_\phi = 200 \times 420 \times 64$ and minimum spacings $\Delta_r = \Delta_\theta = 5.0\times 10^{-4}$, we built two additional grids by doubling either the thickness of the very first layer of cells covering the sphere or the length of the cell closest to the upper pole of the sphere. This yielded grid II with $\Delta_r = 1.0\times 10^{-3}, \Delta_\theta = 5.0\times 10^{-4}$, and grid III with $\Delta_r = 5.0\times 10^{-4}, \Delta_\theta = 1.0\times 10^{-3}$. We also considered a grid refined twice in the radial direction, grid IV, with $\Delta_r = 2.5\times 10^{-4}, \Delta_\theta = 5.0\times 10^{-4}$. The evolution of the lift coefficient on these four grids is shown in figure~\ref{fig6-3}. The largest differences on  $C_L(t)$ (panel $(b)$) are observed with grid II, suggesting that increasing $\Delta_r$ beyond the value selected in grid I is detrimental. In contrast, reducing $\Delta_r$ as in grid IV does not induce any significant change in the results. This is why the discretization based on $\Delta_r = \Delta_\theta = 5.0\times 10^{-4}$ was used throughout this study. 

\section{Influence of artificial perturbations on a marginally stable configuration}
\label{appB}
\begin{figure}
\centering
\includegraphics[width=0.96\textwidth]{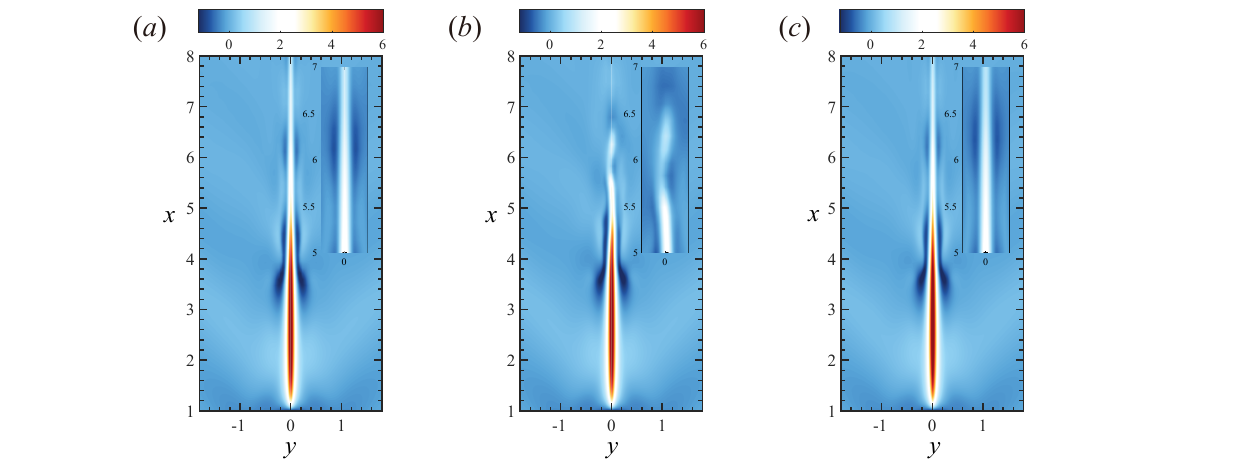}
\caption{Influence of an artificial perturbation on the jet stability in the stable case $(Fr, Re, Pr) = (0.3,100, 700)$. The perturbation is applied in the form $u_x' = 10^{-2} \exp\left\{-400[(y - 0.005)^2 + (x - 2)^2]\right\}$ during the time interval $20 < t < 23$. Contours depict the absolute vertical velocity $u_x - 1$, with insets showing zoomed views of the jet tail. ($a)$: prior to the introduction of the perturbation ($t = 20$); $(b)$: during the application of the perturbation ($t = 22$); $(c)$: long after the perturbation has been removed ($t = 62$). 
}
\label{fig6-4}
\end{figure}

As discussed in \S\,\ref{2.2}, in most cases the jet instability is only triggered by the accumulation of numerical round-off errors. However, in regimes close to the onset of the instability, the natural growth rate of perturbations becomes exceedingly slow. In such cases, we introduce a calibrated external perturbation to expedite the growth of the unstable mode and reduce the computational time required to reach saturation. It is of course important to verify that the imposed disturbances do not fundamentally alter the inherent stability of the flow by inducing spurious instabilities.

To this end, we considered the case $(Fr,Re, Pr) = (0.3,100, 700)$, lying near the threshold of the instability as figure \ref{fig5-19}$(a)$ indicates. With no external forcing, the jet remains stable and retains a perfectly axisymmetric configuration, as depicted in figure~\ref{fig6-4}$(a)$. We then introduced a relatively large artificial perturbation in the form $u_x' = 10^{-2} \exp\left\{-400[(y - y_0)^2 + (x - x_0)^2] \right\}$ localized around $x_0 = 2$ and $y_0 = 0.005$. This perturbation rapidly destabilizes the jet, leading to the development of a pronounced meandering motion during approximately 3000 time steps ($t \approx 43$), as shown in figure~\ref{fig6-4}$(b)$. Then, we removed the perturbation and continued to monitor the evolution of the jet. As figure~\ref{fig6-4}$(c)$ indicates, the flow gradually returns to its original axisymmetric state, confirming that the system is inherently stabie. This return to equilibrium supports the view that the perturbations willingly introduced in the computations, which are two orders of magnitude smaller than that in the above example, do not alter the intrinsic stability of the flow. 

\bibliographystyle{jfm}
\bibliography{main}

\end{document}